\documentclass[11pt]{article} 

\usepackage[ansinew]{inputenc}
\usepackage{amsmath, amssymb, graphics, amsthm}
\usepackage[pdftex]{graphicx, color}
\usepackage{epsfig}
\usepackage{color}
\usepackage{undertilde}
\usepackage{fancyhdr}

\newtheorem*{theorem}{Theorem}

\newcommand{\m}{\mbox{}}

\newcommand{\be}{\begin{equation}}
\newcommand{\ee}{\end{equation}}
\newcommand{\ba}{\begin{eqnarray}}
\newcommand{\ea}{\end{eqnarray}}

\title{{\sf New Variables for Classical and Quantum Gravity}\\
{\sf in all Dimensions II. Lagrangian Analysis}} 
\author{
{\sf N. Bodendorfer}$^1$\thanks{{\sf 
norbert.bodendorfer@gravity.fau.de}},
{\sf T. Thiemann}$^{1,2}$\thanks{{\sf 
thomas.thiemann@gravity.fau.de,
tthiemann@perimeterinstitute.ca}},
{\sf A. Thurn}$^1$\thanks{{\sf 
andreas.thurn@gravity.fau.de}}\\
\\
{\sf $^1$ Inst. for Theoretical Physics III, FAU Erlangen -- N\"urnberg,}\\
{\sf Staudtstr. 7, 91058 Erlangen, Germany}\\
\\
{\sf and}\\
\\
{\sf $^2$ Perimeter Institute for Theoretical Physics,}\\ 
{\sf 31 Caroline Street N, Waterloo, ON N2L 2Y5, Canada}
}
\date{{\small\sf \today}}

\makeatletter
\@addtoreset{equation}{section}
\makeatother
\renewcommand{\theequation}{\thesection.\arabic{equation}}

\begin{document} 

\maketitle

{\sf
\begin{abstract}
We rederive the results of our companion paper, for matching spacetime and internal signature, 
by applying in detail 
the Dirac algorithm to the Palatini action. While the constraint set of the Palatini action contains second class constraints, by an appeal to the 
method of gauge unfixing, we map the second class system to an equivalent  first class 
system which turns out to be identical to the first class constraint system
obtained via the extension of the ADM phase space performed in our companion paper.
Central to our analysis is again the appropriate treatment of the simplicity constraint.
Remarkably, the simplicity constraint invariant extension of the Hamiltonian constraint,
that is a necessary step in the gauge unfixing procedure, involves a correction term which is 
precisely the one found in the companion paper and which makes sure that the 
Hamiltonian constraint derived from the Palatini Lagrangian coincides with the 
ADM Hamiltonian constraint when Gau{\ss} and simplicity constraints are satisfied.       
We therefore have rederived our new connection formulation of General Relativity 
from an independent starting point, thus confirming the consistency of this framework.
\end{abstract}
}

\newpage

\tableofcontents

\newpage

\section{Introduction}

In our companion paper \cite{BTTI}, we developed a higher dimensional connection formulation 
for General Relativity with only first class constraints and Poisson commuting connections in 
any spacetime dimension $D+1\ge 3$. The motivation was to make contact to approaches 
to quantum gravity other than Loop Quantum Gravity \cite{RovelliQuantumGravity, ThiemannModernCanonicalQuantum} which require higher 
dimensions. In our companion paper we discovered this connection formulation by a 
judicious extension of the $D+1$ ADM phase space supplemented by first class Gau{\ss} and simplicity constraints which by construction lead back to the ADM phase space upon
symplectic reduction.

This approach has the advantage that it is rather simple, allows for SO$(1,D)$ or SO$(D+1)$
as structure group irrespective of the spacetime signature and that in addition it admits 
a free parameter that is similar to but yet rather different from the Immirzi parameter in 
$3+1$ dimensions. However, one may ask whether this connection formulation 
can be obtained from an action principle, just as the LQG connection formulation can be 
obtained from the Holst action \cite{HolstBarberosHamiltonianDerived}. In this paper, we answer this question in the affirmative.
The appropriate action to choose will be simply the $D+1$ Palatini action. However, this 
action based approach has its limitations as compared to the Hamiltonian approach \cite{BTTI}:
There is no Immirzi like freedom and the structure group is tied to the spacetime signature:
It is necessarily SO$(1,D)$ for Lorentzian spacetime signature and SO$(D+1)$ for Euclidean 
spacetime signature. This makes this approach less favourable with an eye towards quantisation
of the Lorentzian theory which requires a compact structure group. Yet the efforts of the 
present paper are not in vain as our results confirm the achievements of \cite{BTTI} via an 
alternative route. Maybe the most astonishing outcome is that we obtain a pure first class 
theory while it is well known that the Palatini formulation is plagued by second class constraints.
The resolution of the apparent contradiction is that we have to apply an additional step
in order to arrive at the first class formulation which goes by the name {\it gauge unfixing}. 

In more detail, we do the following: We start from the Palatini formulation of, say, Lorentzian GR in 
$D+1$ spacetime dimensions with structure group SO$(1,D)$. Following strictly Dirac's canonical 
analysis, this formulation, naturally leads to an SO$(1,D)$ connection $A$ and a so$(1,D)$ valued
vector density $\pi$ which is canonically conjugate to the connection. However, in addition
to the SO$(1,D)$ Gau{\ss} constraint, the $D$-dimensional spatial diffeomorphism constraint and the Hamiltonian
constraint, there is an additional primary constraint $S$ which requires the  momentum $\pi$
to derive from (the pull back to the leaves of the foliation of) a co-$(D+1)$-bein. We call it 
simplicity constraint because it is precisely the temporal spatial part of the simplicity constraint
of a higher dimensional Plebanski formulation \cite{FreidelBFDescriptionOf}. The stability of the constraint
$S$ with respect to the canonical Hamiltonian enforces a secondary constraint $D$ and 
$(S,D)$ form a second class pair. The situation is of course completely the same as in $D=3$
dimensions. In $D=3$ dimensions one can now either consider this SO$(1,3)$ connection formulation
and try to quantise the corresponding Dirac bracket \cite{AlexandrovSU(2)LoopQuantum} with non Dirac bracket commuting connections or one imposes the 
time gauge and reduces the (Holst modified) theory to a Dirac bracket commuting SU$(2)$ (or SO$(3)$)
connection formulation. In 
higher dimensions also both possibilities exist except that imposing the time gauge does not
lead to a SO$(D)$ connection formulation but rather the extended ADM formulation already derived
in \cite{HanHamiltonianAnalysisOf}. Thus the second strategy does not lead to the desired connection formulation with compact SO$(D)$ precisely due to the dimensional mismatch between $D$ and 
$D(D-1)/2$. Thus, in order to have a connection formulation only the first possibility remains
but then the complication with the Dirac bracket arises. It is at this point where gauge unfixing
comes into play: By a systematic, allowed  modification of the Hamiltonian constraint which does not alter 
its first class character, the $S$ constraints become Poisson commuting with all but
the $D$ constraints. One can therefore consider the $D$ constraints as gauge fixing conditions for the $S$ constraints and impose only the first class constraints. This way
one can map the second class 
constraint system to an equivalent first class constraint system and replace the complicated 
Dirac bracket by the simple ordinary Poisson bracket with Poisson commuting connections. 
In the end, this formulation is identical to the one of our companion paper for 
matching spacetime and internal signature as well as unit Immirzi like parameter.\\
\\
This paper is is organised as follows:\\ 
\\
In section \ref{section_ham1}, we will perform a canonical analysis of the higher dimensional Palatini theory in the spirit of \cite{PeldanActionsForGravity} following strictly Dirac's procedure \cite{DiracLecturesOnQuantum}. This approach uses vielbeins at every stage of the construction. In 
\ref{section_ham2} we start from a different formulation of the simplicity constraints which 
avoids the introduction of vielbeins and is closer in spirit to the Plebanski formulation of 
General Relativity.  As in the $3+1$-dimensional case, the Hamiltonian formulation of higher dimensional Palatini theory has second class constraints which can either be solved to obtain geometrodynamics or the Dirac bracket has to be implemented on a Hilbert space. To circumvent these problems, we apply the procedure of gauge unfixing which we 
review in in section \ref{section_gu} and apply to GR in section \ref{section_gugr}.
The result is 
an SO$(1,D)$ or SO$(D)$ connection formulation for Lorentzian or Euclidean General
Relativity respectively with first class constraints only and a connection variable which is Poisson self-commuting, the price to pay is one extra term in the Hamiltonian constraint. We finish with some concluding remarks in section \ref{section_cr}.

In the appendix we complete the paper by specialising to $D+1=4$ and supplying the strict 
Immirzi term that is peculiar to $4$ spacetime dimensions.

\section{The $(D+1)$-dimensional Palatini Hamiltonian}

\label{section_ham1}

In this chapter, we will derive the Hamiltonian formulation of $(D+1)$-dimensional Palatini theory. To account for the mismatch of degrees of freedom between the connection and the vielbein, a new constraint, the simplicity constraint, will appear. The Dirac algorithm will be applied to check for further constraints. Finally, we will select a maximal first class set of constraints and solve the second class constraints.

\label{chapter_higherdim}

\subsection{Legendre Transformation}

We start with the Palatini action in $D+1$ dimensions and $\kappa = 1$:
\be S=\frac{s}{2} \int_{\mathcal{M}} d^{D+1}X \, e e^{\mu I} e^{\nu J} F_{\mu \nu IJ} \text{.}\ee 
$\mathcal{M}$ denotes the space-time manifold of topology $\mathbb{R} \times \sigma$. $\mathcal{M}$ foliates into hypersurfaces $\Sigma_t := X_t(\sigma)$, where $X_t : \sigma \rightarrow M$ is an embedding defined by $X_t(\sigma) = X(t, \sigma)$. $x^a\text{,} \, a,b,c = 1, \ldots, D$ are local coordinates on $\sigma$ \cite{ThiemannModernCanonicalQuantum}. $s$ denotes the space-time metric signature, i.e. $g_{\mu \nu} = \text{diag}(s, 1, \ldots, 1)$ for a flat space-time ($\mu, \nu = 0,1,...,D$). Since we start from the Palatini action, Sylvesters theorem tells us that the signature of the internal metric has to coincide with the space-time metric signature. $e^{\mu I}$ denotes the vielbein and $F_{\mu \nu IJ} := \partial_\mu A_{\nu IJ} - \partial_\nu A_{\mu IJ} + [A_\mu, A_\nu]_{IJ}$ is the field strength of the SO$(1,D)$ (SO$(D+1)$) connection $A_{\mu IJ}$, $I,J = 0,...,D$.

The split in space and time is performed analogously to the $4d$ case. We split the time-evolution vector field into
\be  T^\mu = N n^\mu + N^\mu \text{,} ~~ n_\mu N^\mu = 0 \text{.} \ee 
$N$ is called the lapse function and $N^\mu$ the shift vector field. $n^\mu$ is the unit future pointing vector field normal to the spatial slices $\Sigma_t$, i.e. $n^\mu n_\mu = s$ and $n^\mu \partial_\mu X_t = 0$. We use
\be  \delta^\mu \m_\nu= (g^\mu \m_\nu -s n^\mu n_\nu) +s n^\mu n_\nu  =: q^{\mu}\m_{\nu} + s n^\mu n_\nu \ee 
to project the vielbein as
\be  e^\mu_I = q^{\mu}\m_{\nu}\; e^\nu_I +s e^\nu_I n_\nu n^\mu =: \m^\| e^\mu_I + s n_I n^\mu \text{.}\ee 
With $e = N \sqrt{q}$, the action becomes 
\be S= s\int_{\mathcal{M}} d^{D+1}X \, N \left( \frac{1}{2}  \sqrt{q} \m^\| e^{\mu I} \m^\| e^{\nu J} F_{\mu \nu IJ} + s \sqrt{q} n^I n^\mu \m^\| e^{\nu J} F_{\mu \nu IJ} \right) \text{.} \ee 
The next step is to rewrite the action using
\begin{eqnarray}
   N \sqrt{q}  \m^\| e^{\mu I} n^J n^\nu F_{\mu \nu IJ} &=& -\frac{1}{2} \pi'^{\mu IJ} (T^\nu - N^\nu) F_{\mu \nu IJ} \nonumber \\
   &=& \frac{1}{2} \pi'^{\mu IJ} \mathcal{L}_T A_{\mu IJ} + \frac{1}{2} (T^\nu A_{\nu IJ}) G'^{IJ} - N^\mu \mathcal{H}'_\mu 
\end{eqnarray}
and
\be \frac{1}{2} N \sqrt{q} \m^\| e^{\mu I} \m^\| e^{\nu J} F_{\mu \nu IJ} = -s \utilde{N} \mathcal{H}' \text{,} \ee 
where 
\be  \pi'^{\mu IJ} :=  2  n^{[I}  \m^\| E^{\mu |J]} := 2 \sqrt{q} n^{[I}  \m^\| e^{\mu |J]} ~,~~ \utilde{N} := -N / \sqrt{q} ~\text{,} \ee 
\begin{eqnarray} && G'^{IJ} := D_\mu \pi'^{\mu IJ}:=\partial_\mu \pi'^{\mu IJ} + [A_\mu, \pi'^\mu]^{IJ}\text{,} \\ && \mathcal{H}':=\frac{1}{2} \pi'^{\mu IK} \pi'^{\nu J} \m_K F_{\mu \nu IJ} ~~~  \text{and} ~~~ \mathcal{H}'_\mu  := \frac{1}{2} \pi'^{\nu IJ}F_{\mu \nu IJ} \text{.}
\end{eqnarray}  
We use the primes to indicate the dependence on $\pi'^{aIJ}$ and remark that later in the \mbox{analysis}, we will substitute $\pi'^{aIJ}$ by $\pi^{aIJ}$ which denotes the canonically conjugate momentum of $A_{aIJ}$, and neglect the primes. 
The split is completed by pulling back all spatially projected quantities to $\sigma$ by $X$. The connection is pulled back as
\be  A_{aIJ}(t,x) = (\m^\| A_{\mu IJ} \partial_a X^\mu) (X(x,t)) = \m^\|  A_{\mu IJ} (X(t,x)) \partial_a X^\mu (t,x) \text{.} \ee 
Special care has to be taken for the co-$D$-bein, since it has a contravariant external index. We can use the spatially restricted metric $q_{\mu \nu}$ from the ADM formalism to pull the external index down. The pull-back is then possible as
\be  e_a^I(t,x) = (q_{\mu \nu} \m^\| e^{\nu I}  \partial_a X^\mu) (X(x,t)) = q_{\mu \nu} \m^\| e^{\nu I}(X(t,x)) \partial_a X^\mu (t,x)  \text{.} \ee 
Furthermore, we can pull back the restricted metric similar to the ADM formalism, see \cite{ThiemannModernCanonicalQuantum} for details, and use it to raise the spatial index of the pulled back $D$-bein. A push-forward using $X$ of the spatial co-$D$-bein yields again the projected co-$D$-bein we started with when carefully writing out all the terms. 

We split $d^{D+1}X$ into $dt \,d^{D}x$ with $d^{D}x$ being the integration over the spatial coordinates $x^a$ on $\sigma$. Note that $e = N \sqrt{q}$ and $\sqrt{q} d^Dx$ is the induced measure on $\sigma$. 

During the split, the lapse function underwent minor changes but we will still, as in the $(3+1)$-dimensional case, call $\utilde{N} = -N/\sqrt{q}$ the lapse function (of density weight -$1$) and denote it by $N$. As will be shown later, the constraints $\mathcal{H}$, $\mathcal{H}_a$ and $G^{IJ}$ retain their interpretation also in the higher-dimensional case. We will therefore refer to $\mathcal{H}$ as the Hamiltonian constraint, to $\mathcal{H}_a$ as the diffeomorphism constraint and $G^{IJ}$ as the Gau{\ss}  constraint.  

We stress that we have to use the Lie derivative along the time evolution vector field as a time derivative and denote it in this chapter by the standard upper dot. In a special coordinate system where $T^\mu=\text{const.}$, the Lie derivative would reduce to a normal derivative, but in the general case, we have to take the deformation of the spatial slices into account.  

After decomposition as well as the substitution $\lambda_{IJ} = -(T \cdot A)_{ IJ}$ and the abuse of notation $\utilde{N} \rightarrow N$, the action reads
\be  S = \int dt \, L =\int dt \int_\sigma d^{D}x \, \left( \frac{1}{2} \pi'^{aIJ} \mathcal{L}_T A_{aIJ} - N \mathcal{H}' - N^{a} \mathcal{H}'_a - \frac{1}{2} \lambda_{ IJ} G'^{IJ} \right) \text{.} \ee 

The Legendre transformation is performed according to Dirac \cite{DiracLecturesOnQuantum}. The total Hamiltonian is defined by
 \begin{eqnarray}
 H_T &=& \left(\int_\sigma d^Dx \sum_i p_i(x) \dot{q}^i(x) \right)- L + \text{constraints} \nonumber \\
   &= & \int_\sigma d^Dx \biggl( N \mathcal{H}' +  N^{a} \mathcal{H}'_a + \frac{1}{2} \lambda_{ IJ} G'^{IJ} + \alpha P_N + \alpha^a P_{N^a} + \frac{1}{2} \alpha_{IJ} P_{\lambda_{IJ}} \nonumber \\ 
   & &  ~~~~~~~~~~~~~~~~~~~~~~~~~~~~~~~~~~~~~~~~~~~~~  + \frac{1}{2} c_{aIJ} S^{aIJ} + \gamma^a_I P_{E^a_I} \biggr) \text{.} ~~~
\end{eqnarray}

The $P$s denote the canonically conjugate momenta to the indicated variables and $\pi^{aIJ}$ is the momentum to $A_{aIJ}$. We remark that we put a factor of $1/2$ in front of the added terms which are summed over antisymmetric indices because all independent values are summed twice. 
\be  S^{aIJ} = \pi^{aIJ} - \pi'^{aIJ} \ee 
is called the simplicity constraint. We redefine\footnote{
Recall that Lagrange multipliers which depend on phase space variables do not pose any problem during the canonical analysis because when Poisson commuted with other constraints, additional terms appearing due to this dependency are always proportional to constraints.} 
 the Lagrange multiplier of the simplicity constraint to replace all $\pi'^{aIJ}$ by $\pi^{aIJ}$ in $\mathcal{H}'$, $\mathcal{H}'_a$ and $G'^{IJ}$ and drop the primes to indicate the dependence on $\pi^{aIJ}$ only in
$\mathcal{H}$, $\mathcal{H}_a$ and $G^{IJ}$. 

It is convenient to split the simplicity constraint into a boost and a non-boost part by decomposing its multiplier as
\be  c_{aIJ} = 2 n_{[I} \bar{c}_{a|J]} + \bar{c}_{aIJ} \text{.}\ee 
Using the projector $\bar{\eta}^I \m_J := \eta^I \m_J -s n^In_J$, we can explicitly write $\bar{c}_{aIJ} = \bar{\eta}^K \m_I \bar{\eta}^L \m_J c_{aKL}$ and $\bar{c}_{aI} = -s c_{aIJ} n^J$.  In the following, we will use this notation to split other objects with the above index structure in the same way. The simplicity constraint then splits into 
 \begin{eqnarray}  
 \frac{1}{2} c_{aIJ} S^{aIJ} = 0 ~~ & \Leftrightarrow &  \frac{1}{2} \bar{c}_{aIJ} \bar{S}^{aIJ}:= \frac{1}{2} \bar{c}_{aIJ} \bar{\pi}^{aIJ} = 0 \text{,} \nonumber \\
 & & s \bar{c}_{aI} S^{aI} := - \bar{c}_{aI} \left( \pi^{aIJ} n_J + s\;E^{aI} \right) =0 \text{,} 
 \end{eqnarray} 
where the first constraint ensures that $\bar{\pi}^{aIJ} = 0$ and hence $\pi^{aIJ} = 2 n^{[I} B^{a|J]} $ for some $B^{aJ}$ and the second constraint sets $B^{aJ} = E^{aJ}$. We will absorb the factor $s$ in the second line into $\bar{c}_{aI}$ for convenience. 

Notice that the $D+1$ conditions $n^I E^a_I=0,\;n^I n_I=s$ completely fix $n_I$ as a function
of $E^a_I$. It will turn out to be convenient to treat
$n^I$ is as an independent field and to constrain it in the following such that it is given as a function of the $E^{aI}$ on the corresponding constraint surface. 
We therefore add the constraints $n^I n_I-s \approx 0$ and $E^{aI}n_I \approx 0$ to the Hamiltonian with the Lagrange multiplier $\rho$ and $\rho_a$ as well as the primary constraint
that its conjugate momentum vanishes.
After these considerations, the total Hamiltonian reads
\begin{eqnarray}
H_T & = & \int_\sigma d^Dx \biggl( N \mathcal{H} +  N^{a} \mathcal{H}_a  + \frac{1}{2} \lambda _{ IJ} G^{IJ} + \alpha P_N + \alpha^a P_{N^a} + \frac{1}{2} \alpha_{IJ} P_{\lambda_{IJ}} +\frac{1}{2} \bar{c}_{aIJ} \bar{S}^{aIJ} \nonumber\\ 
& & ~~~~~~~~~~~  + \bar{c}_{aI}S^{aI} + \gamma^a_I P_{E^a_I} + \gamma_I P_{n_I} +\rho (n^I n_I -s) + \rho_a E^{aI}n_I \biggr) ~~~~~~~~~ \label{eq:TotalHamiltonian}
\end{eqnarray}
and the symplectic potential is given by
\be  \frac{1}{2} \pi^{aIJ} \dot{A}_{aIJ} + P_{E^a_I} \dot{E}^a_I + P_{n_I} \dot{n}_I + P_N \dot{N} + P_{N^a} \dot{N}^a + \frac{1}{2} P_{\lambda_{IJ}} \dot{\lambda}_{IJ} \text{,}\ee
The non-vanishing Poisson-brackets are accordingly
\begin{alignat}{4}
 \left\{ A_{aIJ}(x), \pi^{bKL}(y) \right\} &= 2 \delta^{(D)}(x-y) \delta_a^b  \eta_I^{[K} \eta _J^{L]}  \text{,} ~~~~& \left\{ E^a_I(x), P_{E^b_J}(y) \right\} &=  \delta^{(D)}(x-y) \delta^a_b \eta^J_I \text{,} 
\\
 \left\{ n_I(x), P_{n_J}(y) \right\} &=  \delta^{(D)}(x-y) \eta^J_I \text{,} ~~~~&
\left\{ N(x), P_{N}(y) \right\} &=  \delta^{(D)}(x-y) \text{,}   \nonumber
\\
 \left\{ N^a(x), P_{N^b}(y) \right\} &=  \delta^{(D)}(x-y) \delta^a_b \text{,} ~~~~& \left\{ \lambda_{IJ}(x), P_{\lambda_{KL}}(y) \right\} &= 2 \delta^{(D)}(x-y) \eta_I^{[K} \eta _J^{L]} \text{.} \nonumber
\end{alignat}

\subsection{Constraint Analysis}

We start the constraint analysis with the total Hamiltonian $H_T$ given in (\ref{eq:TotalHamiltonian}). We need to require preservation of the smeared constraints in time and therefore check upon the weak vanishing of their Poisson brackets with the Hamiltonian $H_T$.  We will denote the 
smearing test functions by $f^{.}_{.}$ with corresponding index structure in order to to distinguish them from the smearing Lagrange multipliers $\alpha,\; \alpha^a,\;\alpha_{IJ},\;\bar{c}_{aIJ},\;\bar{c}_{aI},
\; \gamma^a_I,\;\gamma_I,\;\rho,\;\rho_a$ and smearing phase space functions $N,\; N^a,\; \lambda_{IJ}$   that appear
within $H_T$ and which, in contrast to the test functions, will become in part successively constrained during the stability analysis. The (weak) vanishing of the Poisson bracket of the smeared constraints with the Hamiltonian must hold for all test functions and their full index range.

Since $N$, $N^a$ and  $\lambda_{IJ}$ only appear linearly and their momenta are constrained to vanish, stability of these momenta yields immediately the constraints
\be \mathcal{H} \approx 0~, ~~ \mathcal{H}_a \approx 0 ~,~~ \text{and} ~ ~ G^{IJ} \approx 0  \text{,} \ee
called the Hamiltonian, diffeomorphism, and Gau{\ss} constraints. Stability of $S^{aI}$ requires 
\begin{eqnarray}
  0 & \approx & \int_\sigma d^Dx \, \bar{f}_{aI}(x) \left\{\pi^{aIJ}(x)n_J(x) + s E^{aI}(x), H_T \right\} \nonumber \\
    & = & \int_\sigma d^Dx \, \bar{f}_{aI} \left( *^{aIJ}n_J + \pi^{aIJ} \gamma_J  +s \gamma^{aI}  \right) 
\nonumber \\
    & = & \int_\sigma d^Dx \, \bar{f}_{aI} \left( *^{aIJ}n_J + \bar{\pi}^{aIJ} \gamma_J  +s \bar{\gamma}^{aI}  \right) \text{.} \end{eqnarray} 
In the last line we noticed that only the piece $\bar{\gamma}^{aI}$ orthogonal to $n^I$ survives
since $\bar{f}_{aI}$ is projected. This equation can therefore be solved for 
the Lagrange multiplier $\bar{\gamma}^{aI}$. $*^{aIJ}$ denotes the terms which come from Poisson commuting $\pi^{aIJ}$. We do not compute them explicitly as they will be of no importance after solving the second class constraints. Likewise
\begin{eqnarray}
  0 & \approx & \int_\sigma d^Dx \, f_a(x) \left\{E^{a}_I(x) n^I(x), H_T \right\} \nonumber \\
    & = & \int_\sigma d^Dx \, f_a \left( \gamma^a_I n^I+\gamma^I E^a_I  \right) \text{,}
 \end{eqnarray}  
which can be solved for the piece $\gamma^a_I n^I$ of $\gamma^a_I$.

Next, we calculate 
\be  0\approx \int_\sigma d^Dx \, f(x) \left\{n^I(x)n_I(x)-s, H_T \right\} = 2 \int_\sigma d^Dx \, f \, n^I \gamma_I \text{.} \ee 
This requires that $\gamma^I = \bar{\gamma}^I$, i.e. the Hamiltonian flow does not change the length of  $n^I$. In the same way, 
\be  0 \approx  \int_\sigma d^Dx f^a_I(x) \left\{P_{E^a_I}(x), H_T \right\}  =  \int_\sigma d^Dx  f^{aI} (\bar{c}_{aI}-\rho_a n_I) \text{.} \ee 
It follows that $\bar{c}_{aI}= 0 = \rho_a$. As a consequence, both $\bar{c}_{aI}S^{aI}$ and $\rho_a E^{aI}n_I$ drop from the Hamiltonian. Further, 
\be  0 \approx  \int_\sigma d^Dx \, f_I(x) \left\{P_{n_I}(x), H_T \right\}  \approx  \int_\sigma d^Dx \, f^{I} \left( -2 n_I \rho + \bar{c}_{aIJ} E^{aJ}  \right) \text{.} \ee 
After projecting $f^I$ orthogonally and perpendicularly to $n^I$ we conclude that $\rho = 0$ and $\bar{c}_{aIJ} E^{aJ} = 0$. This means that $\bar{c}_{aIJ}$ is trace free and we can parametrise 
its unconstrained part by a matrix $\bar{c}'_{aIJ}$ orthogonal to $n^I$ with the same symmetries according to 
\be \bar{c}_{aIJ} \equiv \bar{c}^T_{aIJ}=\bar{c}'_{aIJ} - \frac{2}{D-1} E_{a[I} \bar{c}'_{cK|J]} E^{cK} \text{.} \label{eqn_traceprojector} \ee
Here 
$E_{aI}$ denotes the inverse of $E^{aI}$ defined by $E^{aI} E_{bI} = \delta^a_b$ and $E_{bI}n^I=0$, i.e. $E^{aI} E_{aJ}=\bar{\eta}^I_J$.

 The remaining constraint $\bar{S}^{aIJ}$ yields
\begin{eqnarray} \label{trace}
 0 &\approx&  \int_\sigma d^Dx \, \bar{f}_{aIJ}(x) \left\{\bar{\pi}^{aIJ}(x), H_T \right\} \nonumber \\ 
 & \approx& \int_\sigma d^Dx \, \bar{f}_{aIJ} \biggl(  2 N  D_b  \left(\pi^{b[I|K} \pi^{a|J]K} \right) \nonumber \\
 & &~~~~~~~~~~~~~~~~~~ + 2 E^{a[J}\left(-\bar{\gamma}^{I]} - n_K \lambda^{K|I]} +s (\partial_b N)E^{b|I]} +N^b\; D_b n^{I]} \right)  \biggr) 
\nonumber \\ 
& \approx & \int_\sigma d^Dx \,  \biggl[ 2s N \bar{f}_{aIJ} E^{b[I} D_b E^{a|J]} \,  
\nonumber\\
 & &~~~~~~~~~~~~~~~~~~ + 2 \bar{f}_{aIJ} E^{a[J}\left(-\bar{\gamma}^{I]} - n_K \lambda^{K|I]} +s (\partial_b N)E^{b|I]} + N^b\; D_b n^{I]} \right)  \biggr) \biggr] \text{,}
\end{eqnarray}
where $D$ is the covariant differential associated with the connection $A$ (it acts only
on internal indices and neglects tensorial indices and density weights). We have used the 
simplicity constraint several times in order to arrive at this weak identity.
We can decompose the test function as $\bar{f}_{aIJ}=\bar{f}^{\text{TT}}_{aIJ}+2e_{a[I}\bar{f}_{J]}$ and 
conclude that the trace and trace free part of (\ref{trace}) have to vanish separately.
Consider the quantity
\be D^{aIJ}:=2 s E^{b[I}\; D_b E^{aJ]} \text{.} \ee
It is neither transversal nor trace free. Since the test function $\bar{f}_{aIJ}$ is transversal
actually only the transversal part 
\be \bar{D}^{aIJ}=\bar{\eta}^I_K\; \bar{\eta}^J_L\; D^{aKL} \ee
enters (\ref{trace}). We split off the its trace part 
\be \bar{D}^{aIJ}=\bar{D}^{aIJ}_T+\frac{2}{D-1} E^{a[I} E_{bK} \bar{D}^{bK|J]}\ee
and rewrite (\ref{trace}) as 
\ba
0 & \approx & \int_\sigma d^Dx \,   N \bar{f}_{aIJ}\;  \,  
 \nonumber\\
 & &~~~~\times \biggr[ \bar{D}^{aIJ}_T+ 2 E^{a[J}\left(-\bar{\gamma}^{I]} - n_K \lambda^{K|I]} +s (\partial_b N)E^{b|I]} + N^b\; D_b n^{I]}-\frac{1}{D-1} E_{bK} \bar{D}^{bK|I]} \right)  \biggr] 
 \nonumber\\
&=:& \int_\sigma d^Dx \, [\bar{f}^T_{aIJ}  N\; \bar{D}^{aIJ}_T
+2\bar{f}_I\;(\bar{\gamma}^{I}+\frac{N}{D-1} \bar{D}^{I})] \text{.}
\end{eqnarray}
Vanishing of the trace part now requires that $\bar{\gamma}_I=-\frac{N}{D-1} \bar{D}_I$ and vanishing
of the trace free part yields the secondary constraint $\bar{D}^{aIJ}_T$ which is manifestly 
transversal and trace free.\\
\\
To summarise, the outcome of the analysis so far is that 
$\bar{c}_{aI}=\rho=\rho_a=\gamma^I n_I=\bar{c}_{aIJ} E^{aJ}=0$ and that 
$\gamma^a_I=\gamma^a_{I0},\;\bar{\gamma}^I=\bar{\gamma}^I_0$ are fixed functions on phase
space. The Hamiltonian that stabilises the primary constraints therefore reduces to 
\begin{eqnarray}
 H_T &=& \int_\sigma d^Dx \biggl( N \mathcal{H} +  N^{a} \mathcal{H}_a  +  \frac{1}{2} \lambda _{ IJ} G^{IJ} + \alpha P_N + \alpha^a P_{N^a} + \frac{1}{2} \alpha_{IJ} P_{\lambda_{IJ}} + \frac{1}{2} \bar{c}^T_{aIJ} \bar{S}^{aIJ} \nonumber \\ 
 & & ~~~~~~~~~~~~   + \gamma^a_{I0} P_{E^a_I} + \bar{\gamma}_{I0} P_{n_I} 
 \biggr)
\end{eqnarray} 
and the secondary constraints are $G^{IJ},\;\mathcal{H}_a,\; \mathcal{H},\; \bar{D}^{aIJ}_T$.  Here
$\bar{c}^T_{aIJ}$ denotes the trace free part of  $\bar{c}_{aIJ}$ etc.\\
\\
In the next step of the canonical analysis, the secondary constraints have to be checked for consistency and then we must subdivide (linear combinations of) the constraints into those of 
first and second class respectively. Anticipating the result, we will show that $S,D$ form a 
second class pair while linear combinations of the remaining constraints are first class. 
We begin with the constraints $\mathcal{H}$, $\mathcal{H}_a$, and $G^{IJ}$
which only depend on the variables $\pi^{aIJ}$ and $A_{aIJ}$ but not on 
$N,\;N^a\; \lambda_{IJ},\;n^I,\; E^a_I$ 
whence they have vanishing brackets with the primary constraints 
$P,\; P_a,\; P_{\lambda_{IJ}},\; P_{n^I},\; P_{E^a_I}$. Thus, to ensure their stability, only their 
brackets with themselves and the trace free part of the simplicity constraint $\bar{S}^{aIJ}$
appearing in the Hamiltonian above have to be checked.

We begin with the Gau{\ss} constraint. The action of the smeared Gau{\ss} constraint on the phase space variables is
\begin{eqnarray}
	 \left\{A_{aIJ}, \frac{1}{2} G^{KL}[f_{KL}] \right\} & = & - D_a f_{IJ}\text{,} \\
   \left\{\pi^{aIJ}, \frac{1}{2} G^{KL}[f_{KL}] \right\} & = &  \left[ f, \pi^{a} \right]^{IJ} \text{,} 
\end{eqnarray}
meaning that $A_{aIJ}$ transforms as an internal connection and $\pi^{aIJ}$ as an internal second rank contravariant tensor. The Gau{\ss}  constraint closes with itself, 
\be  \left\{ \frac{1}{2} G^{IJ}[f_{IJ}],  \frac{1}{2} G^{KL}[\gamma_{KL}] \right\}  =  \frac{1}{2} G^{IJ}\left[\lambda_{IK} \gamma^K \m_J -\gamma_{IK} \lambda^K \m_J    \right] \text{,}\ee 
which was expected as the difference between two rotations is again a rotation. The factors $1/2$ are included because of the antisymmetry of the multiplier. Since the Hamiltonian constraint and the diffeomorphism constraint do not have free internal indices, the calculations yield as expected
\begin{eqnarray}
	 \left\{ \frac{1}{2} G^{IJ}[f_{IJ}],  \mathcal{H}_a[N^a] \right\} & = & 0 \text{,}\\
   \left\{ \frac{1}{2} G^{IJ}[f_{IJ}],  \mathcal{H}[N] \right\} & = & 0 \text{.} 
\end{eqnarray}
It follows that
\be \left\{ \frac{1}{2} G^{IJ}[f_{IJ}],  H_T \right\}  \approx - \bar{c}^T_{aIJ} E^{aI} n^K \lambda_{K} \m^J = 0 \text{,} \ee 
since $\bar{c}^T_{aIJ}$ is trace free whence the Gau{\ss} constraint is already stable.

The diffeomorphism constraint as given generates spatial diffeomorphisms mixed with internal rotations. In order to see this, we define
\be  \label{eqn_hprime} \tilde{\mathcal{H}}_a := \mathcal{H}_a - \frac{1}{2} A_{aIJ} G^{IJ} = \frac{1}{2} \pi^{bIJ} \partial_a A_{bIJ}- \frac{1}{2} \partial_b \left(\pi^{bIJ}A_{aIJ} \right) \text{.}\ee 
The action of $\tilde{\mathcal{H}}_a$ on the phase space variables is
\begin{eqnarray}
\left\{ A_{aIJ}, \tilde{\mathcal{H}}_b[f^b]\right\} &=& f^b \partial_b A_{aIJ} + (\partial_a f^b) A_{bIJ}  = \mathcal{L}_f A_{aIJ} \text{,}\\
\left\{ \pi^{aIJ}, \tilde{\mathcal{H}}_b[f^b]\right\} &=& f^b \partial_b \pi^{aIJ} - (\partial_b f^a) \pi^{bIJ} + (\partial_b f^b) \pi^{aIJ} = \mathcal{L}_f \pi^{aIJ}\text{,}
\end{eqnarray}
meaning that $A_{aIJ}$ transforms like the components of a one-form and $\pi^{aIJ}$ like the components of a vector density under infinitesimal diffeomorphisms generated by the Lie derivative.  From this and \eqref{eqn_hprime}, we can deduce
\begin{eqnarray}
\left\{ \mathcal{H}_a[f^a],  \mathcal{H}_b[N^b] \right\} & = &  \mathcal{H}_a[(\mathcal{L}_f N)^a] -\frac{1}{2}  G^{IJ}[f^a N^b F_{abIJ}] \text{,}\\
\left\{ \mathcal{H}_a[f^a],  \mathcal{H}[N] \right\} & = &  \mathcal{H}[\mathcal{L}_f N]  + G^{IJ}[N f^a \pi^{b} \m_I \m^K F_{abJK}] \text{.}
\end{eqnarray}
The Lie derivatives are given by $(\mathcal{L}_f N)^a = f^b \partial_b N^a-N^b \partial_b f^a$ when $N^a$ is considered as a vector field and  $(\mathcal{L}_f N) = f^b \partial_b N-N \partial_b f^b$ where $N$ is considered as a scalar density of weight minus one. 
It follows
\be \left\{ \mathcal{H}_b [f^b],  H_T \right\} \approx \left\{ \tilde{\mathcal{H}}_b [f^b],  H_T \right\}  \approx  N^b \bar{c}^T_{aIJ} E^{aJ}(\partial_b  n^I) =0 \text{,}   \ee 
again due to trace freeness of $\bar{c}^T_{aIJ}$. Thus also the spatial diffeomorphism constraint is already stabilised.

Left with the Hamiltonian constraint, we calculate
\begin{eqnarray} 
\left\{ \mathcal{H}[f],  \mathcal{H}[N] \right\}  &=&   -\frac{1}{2} \mathcal{H}_a \left[(f\partial_bN-N\partial_b f) \pi^{aIJ}\pi^b_{IJ} \right] \nonumber \\ 
& & +\int_\sigma d^Dx \, \frac{3 s}{2} (D-3)! (f \partial_aN-N\partial_a f) \pi^{a}_{[IJ}\pi^b_{KL]}  \pi^{cIJ}F_{cb}^{KL} \nonumber \\
 &\approx&  -\frac{1}{2} \mathcal{H}_a \left[(f \partial_bN-N\partial_b f) \pi^{aIJ}\pi^b_{IJ} \right] \text{,}
\end{eqnarray}
which is more involved than the other calculations
but it is helpful to make use of the fact 
that due to the antisymmetry of the Poisson bracket, only terms proportional to $f\partial_bN-N\partial_bf$ can survive.
 The second term vanishes because of the total antisymmetrisation in the $\pi$s when using the simplicity constraint. 
 We used an important relation concerning the contraction of so$(1,D)$ structure constants which is discussed in appendix \ref{appendix_liealgebra}. 
It follows by the same calculation as in (\ref{trace}) that 
\be 
\{\mathcal{H}[f],H_T\}\approx\{\mathcal{H}[f],\frac{1}{2}\bar{S}^{aIJ}[\bar{c}^T_{aIJ}]\}\approx
-\frac{1}{2} \bar{D}_T^{aIJ}[ f \bar{c}^T_{aIJ}] \approx 0 \text{.}
\ee
In conclusion, the Gau{\ss}, spatial diffeomorphism and Hamiltonian constraint are already
stabilised.\\
\\
Therefore, among the secondary constraints, the only constraint left to be checked for consistency is $\bar{D}_T^{aIJ}$. For this, it will be important to analyse the Poisson bracket
\be  \left\{\bar{S}^{aIJ}[\bar{f}_{aIJ}], D^{bKL}[\bar{f}^{\prime T}_{bKL}]\right\}  = 
\int \; d^Dx\;  \bar{f}_{aIJ} F^{aIJ, bKL} \bar{f}^{\prime T}_{bKL}\ee 
with
\be  \label{tensor} F^{aIJ, bKL}= 4 s  E^{a[K} \bar{\eta}^{L][J} E^{b|I]} 
\ee 
and we have absorbed the transverse trace free projections in $\bar{D}^{aIJ}_T$ into transverse  
trace free smearing functions.
Using this definition, the stability requirement for $D^{aIJ}$ reads
\be  \{D^{bKL}[\bar{f}^T_{bKL}],H_T\}=\int_\sigma d^Dx \,  \bar{f}^T_{aIJ} \left(-\frac{1}{2} F^{aIJ,bKL}\bar{c}^T_{bKL}+ \bar{\Sigma}^{aIJ}_T \right) \approx 0 \text{,} \ee 
where $\bar{\Sigma}^{aIJ}_T$ denotes all the contributions from $H_T$ 
different from $\bar{S}^{aIJ}[\bar{c}^T_{aIJ}]$. We remark that all partial derivatives were integrated away from $\bar{f}^T_{aIJ}$ onto $\bar{\Sigma}^{bKL}_T$ and that a possible longitudinal part of
$\bar{\Sigma}^{aIJ}$ of the form $n^{[I} \bar{\Sigma}^{aJ]}$ and a possible trace part of the 
form $E^{a[I} \bar{v}^{J]}$ would vanish because of the contraction with $\bar{f}^T_{aIJ}$. 

Thus we have to solve 
\be \label{inversion} \frac{1}{2} [F^{aIJ, bKL} \bar{c}^T_{bKL}]_{TT} = \bar{\Sigma}^{aIJ}_T \text{,} \ee 
where $(.)_{TT}$ denotes the transverse trace free part of a 
tensor. Notice that the tensor $F$ in (\ref{tensor}) is symmetric 
under the exchange of the index triples $aIJ$ and $bIJ$, antisymmetric in $IJ$ and $KL$ but 
not  trace free (contraction with $E_{aI}, E_{aJ}, E_{bK}, E_{bL}$ does not vanish) while manifestly transverse (contraction with $n_I, n_J, n_K, n_L$ vanishes). Notice also that the index positions 
of $a,b$ cannot be interchanged as otherwise contraction of $F$ with a trace free tensor
would vanish. The strategy to solve (\ref{inversion})
will be to compute the inverse of $F$ on the space of just transverse tensors and then to determine 
the trace free projection of its action on trace free and transverse tensors. A lengthy but straight forward calculation yields
\be  \left( F^{-1} \right)_{aIJ,bKL} = \frac{s}{4} E_{aA} E_{bB} \left( \bar{\eta}^{AB} 
\bar{\eta}_{K[I} \bar{\eta}_{J]L} - 2 \bar{\eta}^B_{[I} \bar{\eta}_{J][K} \bar{\eta}^A_{L]}\right) \text{.} \ee 
This formula can be discovered by the observation that the two tensors on the right hand 
side are the only traverse tensors with the correct index structure and the same symmetries
as $F$. One then just has to calculate the corresponding contractions in order to determine
the coefficients displayed. Thus 
\be 
F^{aIJ,cMN}\; (F^{-1})_{cMN,bKL}=\delta^a_b\; \bar{\eta}^I_{[K}\; \bar{\eta}^J_{L]}
\ee
is the unit operator on transverse tensors of type $\bar{c}^{bKL}$ which are antisymmetric
in $K,L$. Since $F$ is invertible, we can set 
\be  \bar{c}^{T0}_{aIJ} = 2 P_{aIJ}^{bKL}\; (F^{-1})_{bKL, cMN} \bar{\Sigma}_T^{cMN} \text{,} \ee 
with the transversal trace freeness projector
\be 
P^{aIJ}_{bKL}=\delta^a_b\; \bar{\eta}^I_{[K}\; \bar{\eta}^J_{L]}
+\frac{2}{D-1} E^{a[I}\;\bar{\eta}^{J]}_{[K}\; E_{bL]} \text{.}
\ee
~\\
All constraints are stable at this point and we are left with the Hamiltonian
\begin{eqnarray}
 H_T &=& \int_\sigma d^Dx \biggl( N \mathcal{H} +  N^{a} \mathcal{H}_a  + \frac{1}{2} \lambda _{ IJ} G^{IJ} + \alpha P_N + \alpha^a P_{N^a}  + \frac{1}{2} \alpha_{IJ} P_{\lambda_{IJ}} \nonumber \\
& & ~~~~~~~~~~~ + \frac{1}{2} \bar{c}^{T0}_{aIJ} \bar{S}^{aIJ} + \gamma^a_{I0} P_{E^a_I} + \bar{\gamma}_{I0} P_{n_I}  \biggr) \text{,} ~~~~~~~
\end{eqnarray} 
where the "0" remind us of the fact that those Lagrange multipliers 
have been replaced by phase space dependent functions.

\subsection{Degrees of Freedom}

In this section, we count the degrees of freedom of the derived Hamiltonian system to see if they match those of General Relativity. Before we can count the constraints, we have to identify a maximal subset of first class constraints. Comparing with the other available canonical formulations, we suspect the first class constraints to be $\mathcal{H}$, $\mathcal{H}_a$, $G^{IJ}$, $P_N$, $P_{N^a}$, and $P_{\lambda_{IJ}}$. The first three constraints from this list are not of the first class by themselves. Since the following works for all three constraints, we will denote by $C$ either $\mathcal{H}[N]$, $\mathcal{H}_a[N^a]$, or $G^{IJ}[\lambda_{IJ}]$. We can make all of them first class with respect to the $D$ constraint by the substitution
\be C \rightarrow C - \int \bar{S}^{aIJ}_T \left( \left\{ \bar{D}_T, \bar{S}_T \right\}^{-1} \right)_{aIJ, bKL} \left\{\bar{D}^{bKL}_T, C \right\} \text{.} \ee
This notation is symbolic (notice that the pointwise Poisson brackets are distributional):
The matrix $\{\bar{D}_T,\bar{S}_T\}$ is ultralocal and what is meant is its non distributional
factor.  
As shown earlier in the constraint analysis, $C$ is first class with respect to the traceless part of $\bar{S}^{aIJ}$. In fact, only the Hamiltonian constraint contributes and yields the constraint 
$\bar{D}_T$. The trace and the traceless parts of $\bar{S}^{aIJ}$ and its multiplier are accessible via the projector defined in (\ref{eqn_traceprojector}). Next, we need to check the Poisson bracket with $S^{aI}$. It turns out that the substitution 
\be C \rightarrow C +s\int  \left\{C, \pi^{aIJ} \right\} n_J P_{E^a_I} \ee
forces $C$ to Poisson commute with $S^{aI}$ and all the other constraints. The only constraint which does not Poisson commute with $C$ at this point is the trace part of $\bar{S}^{aIJ}$. Again, we perform a substitution 
\be C \rightarrow C - \frac{s}{D-1} E_{bK} \left\{C, \pi^{bKJ} \right\} E^a_J n_I P_{E^a_I} + \frac{1}{D-1} E_{aI} \left\{ C, \pi^{aIJ} \right\} \bar{\eta}_{JK} P_{n_K} \ee
and see that $C$ is now of the first class. For this, we have to recheck that $C$ still Poisson commutes with $S^{aI}$, which it does. 

The rest of the constraints are of the second class. An easy proof for this statement is to refer to the next section, where the constraints are solved. Since the solution of the proposed second class constraints yields a non-degenerate Poisson bracket, we are sure that we did not solve any first class constraints. It is however possible to explicitly decompose the second class constraints into second class pairs. 

First of all, we note that except for $\bar{D}^{aIJ}_T$, all remaining constraints Poisson commute with the trace free part of $\bar{S}^{aIJ}$. We can thus perform the substitution 
\be C \rightarrow C - \int \bar{S}^{aIJ}_T \left( \left\{ \bar{D}_T, \bar{S}_T\right\}^{-1} \right)_{aIJ, bKL} \left\{D^{bKL}_T, C \right\} \ee
for all the remaining constraints and the first set of second class pairs is
\be  \left\{ \bar{S}^{aIJ}_T [\bar{f}^T_{aIJ}], \bar{D}^{bKL}_T[\bar{g}^T_{bKL}] \right\} =  \int_\sigma d^Dx \, \bar{f}^T_{aIJ} F^{aIJ, bKL} \bar{g}^T_{bKL} \text{.} \ee
It is easy to see that another set of pairs is given by
\be \left\{ P_{E^a_I}[f^a_I], S^{aI}[\bar{g}_{aI}] + E^{aI}n_I[g_a] \right\} = \int_\sigma d^Dx \, 
f^{aI} \left(-s \bar{g}_{aI}-n_I g_a \right) \text{.}\ee
To obtain yet another set of second class pairs which Poisson commute with the first two pairs, we realise that the linear combination
\be P'_{n_I}[\gamma_I] := P_{n_I}[\gamma_I] +s P_{E^a_J} [\gamma_I \pi^{aI} \m_J] \ee
Poisson commutes with all the above constraints. The third set of second class pairs is given by 
\be \left\{ P'_{n_I}[f_I], (n^In_I-s)[g]+\bar{S}^{aIJ}[\bar{g}_{[I} E_{a|J]}] \right\} \approx - \int_\sigma d^Dx \, f_I \left( 2 g n^I -s (1-D) \bar{g}_I \right) \text{.}\ee
We emphasise that the three sets of pairs yield three invertible Dirac matrices. Since constraints from different sets of pairs Poisson commute with each other, the whole Dirac matrix is invertible because its determinant is the product of the three subdeterminants coming from the three sets. 

The transition to the extended Hamiltonian is not necessary since all first class constraints are already contained with arbitrary multipliers in the total Hamiltonian. 

The counting of the degrees of freedom goes as follows: \\ 
\mbox{} \\
\begin{center}
\begin{tabular}[c]{ c | c || c | c  }
  Variable & DoF &  Constraint & Number \\
 \hline
\hline
  $A_{aIJ}$ & $ \frac{D^2(D+1)}{2}$ & First class & (count twice!)\\
  $\pi^{aIJ}$ & $ \frac{D^2(D+1)}{2}$ & $\mathcal{H}$ & $1$ \\
  $\lambda_{IJ}$ & $\frac{D(D+1)}{2}$ & $\mathcal{H}_a$  &$ D$ \\
  $N$ & $1$ & $G^{IJ}$ & $\frac{D(D+1)}{2}$ \\
  $N^a$ & $D$ & $P_N$  & $1$ \\
  $P_{\lambda_{IJ}}$ & $\frac{D(D+1)}{2}$ & $P_{N^a}$ &$D$ \\
  $P_N$ & $1$ & $P_{\lambda_{IJ}}$  &  $\frac{D(D+1)}{2}$  \\  \cline {3-2} \cline{4-2}
  $P_{N^a}$ & $D$ & Second class & \\
  $n^I$ & $D+1$ &  $S^{aIJ}$ & $\frac{D^2(D+1)}{2}$ \\
  $E^{aI}$ & $D(D+1)$ & $\bar{D}^{aIJ}_T$ & $\frac{D^2 (D-1)}{2}-D$ \\
  $P_{n_I}$ & $D+1$ &   $E^{aI}n_I$ &  $D$ \\
  $P_{E^{aI}}$ & $D(D+1)$ & $n^I n_I-s$ & $1$ \\
  &  & $P_{E^{aI}}$ & $D(D+1)$ \\
  &  & $P_{n_I}$ & $D+1$ \\ 
  \hline \hline
  Sum: & $D^3+4D^2+7D+4$ & Sum: & $D^3 + 3D^2 + 8D+6$ 
\end{tabular} 
\end{center}
\mbox{} \\

The difference between the degrees of freedom and the effective number of constraints is $(D-2)(D+1)$ and matches the degrees of freedom of gravity in the Hamiltonian formulation.

\subsection{Solution of the Second Class Constraints}
\label{sec_solution1}

The solution of the second class constraints is done analogously to the treatment by Peldan 
\cite{Peldan1992}. We start by solving the first class constraints $P_N$, $P_{N^a}$, and $P_{\lambda_{AB}}$ by treating $N$, $N^a$, and $\lambda_{AB}$ as Lagrange multipliers. 

To solve the second class constraints, we use the Ansatz
\be A_{aIJ} = \Gamma_{aIJ}+K_{aIJ} \text{.} \ee 
$\Gamma_{aIJ}$ is the hybrid spin connection defined by
\be \nabla_a E^{bI} = \partial_a E^{bI} + \Gamma _{ac}^b E^{cI} + \Gamma_a \m^I \m_J E^{bJ}- \Gamma^c_{ca}E^{bI}=0 \ee
and is explicitly given by
\be \Gamma_{aIJ} = 2s E_{b[I} n_{J]} n^K \partial_a E^b_K + E_{b[I} \bar{\eta}_{J]K} \partial_a E^{bK} + \Gamma_{ab}^c E^b_{[J} E_{c|I]} \text{,} \label{eqn_hybridconnection} \ee
where
\be \Gamma_{ab}^c := \frac{1}{2} q^{cd} \left( \partial_a q_{bd} + \partial_b q_{ad} - \partial_d q_{ab} \right) \ee
is the Levi-Civita connection. 
Notice that
\be \nabla_a E^{aI} = \partial_a E^{aI} + \Gamma_a \m^I \m_J E^{aJ}= 0 \text{.} \ee
Further, we decompose $K_{aIJ}$ into $\bar{K}_{aIJ}$ and $2n_{[I}K_{a|J]}$ similar as for 
$\pi^{aIJ}$ where $K_{aI}\propto K_{aIJ} n^J$ is automatically transversal. At the same time, we solve the simplicity constraints to $\pi^{aIJ} = 2 n^{[I} E^{aJ]}$. The requirement that $n^I$ is orthogonal to $E^{aI}$ and has unit length leads to 
\be  n^I = \frac{\epsilon^{I J_1 \ldots J_D} E^{a_1}_{J_1} \ldots E^{a_1}_{J_D}\epsilon_{a_1 \ldots a_D}}{D!\sqrt{\det E^{aI} E^b_I}} \text{.} \ee 
In the following, we will always mean $n_I = n_I(E^{aJ})$ and thus have solved the constraints $n^In_I-s \approx 0$, $E^{aI}n_I \approx 0$, and $P_{n_I} \approx 0$. 

The boost (longitudinal) part of the Gau{\ss} constraint becomes
\be  n_{[I} \lambda_{J]} D_a \pi^{aIJ} = s \lambda_I \bar{K}_a \m^{KI} E^a_K \label{eq:TraceRotKomp} \ee 
and coincides with the trace part of $\bar{K}_{aIJ}$. Next, we insert this Ansatz into $D^{aIJ}$ and calculate
\begin{eqnarray}
  f^T_{aIJ} \bar{D}^{aIJ}_T &=& \bar{f}^T_{aIJ} D^{aIJ} = -2 s \bar{f}^T_{aIJ} E^{b[I} D_b E^{a|J]} \nonumber \\
                        &=& -2 s \bar{f}^T_{aIJ} E^{b[I} \left( D_b E^{a|J]} + \Gamma_{bc}^a E^{c|J]} - \Gamma^c_{bc} E^{a|J]} \right) \nonumber \\
                        &=& \frac{1}{2} \bar{f}^T_{aIJ} F^{aIJ,bKL}\bar{K}^T_{bKL}  \text{.} 
\end{eqnarray}
In the second line, we have added terms to construct a covariant derivative compatible with $E^{aI}$. Both added terms are zero, the first because of the antisymmetry in $[IJ]$ and the second one because of the trace freeness of $\bar{f}^T_{aIJ}$. We see that $\bar{D}^{aIJ}_T=0$ implies the vanishing of the trace free part of $\bar{K}_{aIJ}$. Thus Gau{\ss} and $D$ constraint together
imply that $\bar{K}_{aIJ}$ vanishes whence $K_{aIJ}=2n_{[I} K_{aJ]} $.

Since we solved second class constraints, we have to perform a symplectic reduction and determine the new symplectic structure. In addition to the above considerations, we set $P_{E^a_I}=0$. The symplectic potential now reads
\begin{eqnarray}  
& &\frac{1}{2} \pi^{aIJ} \dot{A}_{aIJ}  = n^{[I}E^{a|J]} \left( \dot{\Gamma}_{aIJ} + \dot{K}_{aIJ} \right) \nonumber \\
& = &n^I \left((\nabla_a E^a_I)\dot{\m}-\nabla_a \dot{E}^a_I + E^{aJ}\dot{K}_{aIJ} \right) \nonumber \\
& = &-\partial_a(n^I \dot{E}^a_I)+n^I E^{aJ} \dot{K}_{aIJ} \nonumber \\
& = &-\dot{E}^{aJ} n^I K_{aIJ} - \dot{n}^I E^{aJ} K_{aIJ} \nonumber \\
& = & \dot{E}^{aJ}(n_J E^I_a \bar{K}_{bIK}E^{bK}-s K_{aJ}) \nonumber \\
& =: & \dot{E}^{aJ}K'_{aJ}\text{,}
\end{eqnarray}
where we have dropped total time derivatives and divergences, in the second before the 
last step we used that $\dot{n}^I$ is transversal.
\footnote{We also used 
$\nabla_a n^I=0$ which follows from $E^a_I n^I=n^I n_I-s=\nabla_a E^b_I=0$: We have
for the longitudinal part $n_I \nabla_a n^I=\nabla_a (n_I n^I/2)=0$ and for the transversal part
$E^b_I \nabla_a n^I=\nabla_a (E^b_I n^I)=0$.} Notice also that we keep the trace part
of $\bar{K}_{aIJ}$ since we do not solve first class constraints at this point. 

In the last step, we have to express the remaining constraints $\mathcal{H}$, $\mathcal{H}_a$, and $G^{IJ}$ in terms of the new canonical variables. The calculation yields
\begin{eqnarray}
\frac{1}{2} f_{IJ} G^{IJ} &=& -f^{IJ} E^a_{[I} K'_{a|J]} \text{,}\\
N^a \mathcal{H}_a & \approx & 2s N^a  \nabla_{[a} E^ {bJ}   K'_{b]J} \text{,}\\
N \mathcal{H} & \approx & N \left( s \frac{1}{2}E^{aI} E^{bJ} R_{abIJ} - E^{a[I}E^{b|J]} 
K'_{aI} K'_{bJ} \right) \label{reduction_h} \text{.}
\end{eqnarray}
We have neglected terms proportional to the Gau{\ss} constraint in the expressions for $N^a\mathcal{H}_a$ and $N \mathcal{H}$ as well as total derivatives. $R_{abIJ} := 2\partial_{[a} \Gamma_{b]IJ} + [ \Gamma_a, \Gamma_b]_{IJ}$ denotes the field strength of the hybrid spin connection. Up to a global factor, the time gauge $n^I=(1, 0, \ldots, 0)$ and the solution of the boost part of the Gau{\ss} constraint lead to the ADM formalism with internal SO$(D)$ gauge group as derived in \cite{ThiemannModernCanonicalQuantum}.

\section{Equivalent Formulation}
\label{section_ham2}

\subsection{The BF-Simplicity Constraint}

Freidel, Krasnov and Puzio \cite{FreidelBFDescriptionOf} have shown that higher-dimensional Einstein gravity can be written as a constrained BF theory. Since their starting point is a BF theory where there are a priori no vielbeins, they have to use a different simplicity constraint to avoid introducing vielbeins. In the following, we will review their simplicity constraint and discuss its relation to our formulation. To distinguish the two versions of the simplicity constraint, we will call the simplicity constraint dealt with in this section the BF-simplicity. 

We will sketch the main proof given in \cite{FreidelBFDescriptionOf} and adapt it to our canonical treatment. This step is necessary because the original paper considers path integral quantisation for which only the action is needed and the whole vielbein is encoded in an object $B^{\mu \nu}_{IJ}$. In our case, the components $\utilde{N}$ and $N^a$ of the vielbein are not
considered because they function as Lagrange multipliers. In the following, we will denote by $\overline{M}$ a $(D-3)$ multi-index.

Let us begin by citing the main result of \cite{FreidelBFDescriptionOf}:

\begin{theorem}[Freidel, Krasnov, Puzio]  \mbox{} \\ \nonumber
In dimension $D>3$ a generic B field satisfies the constraints
\end{theorem}
\be \epsilon^{\overline{M}IJKL} \tilde{B}^{\mu \nu}_{IJ}\tilde{B}^{\rho \sigma}_{KL} = \tilde{\epsilon}^{[\alpha]\mu \nu\rho\sigma} \tilde{c}^{\overline{M}}_{[\alpha]} \label{eqn_theoremfreidelcondition}\ee
{\it{for some coefficients $c$ with $[\alpha]$, $\overline{M}$ totally skew tensorial and Lie 
algebra combinations respectively of length $D-3$,
if and only if it comes from a frame field. In other words, a non-degenerate $B$ satisfies the constraints (\ref{eqn_theoremfreidelcondition}) if and only if there exist $e^\mu_I$ such that}}
\be \tilde{B}^{\mu \nu}_{IJ} = \pm |e|e_I^{[\mu} e_j^{\nu]} \text{,} \ee
{\it{where $|e|$ is the absolute value of the determinant of the inverse matrix $e_ {\mu I}$.}}
\\

The theorem also holds for $D=3$ with the additional appearance of a topological sector which we will neglect in the following. Clearly, our desired simplicity constraint guaranteeing $\pi^{aAB} = 2 n^{[A} E^{a |B]}$ seems to be a special case of this theorem. 

The constraints are divided into the categories
\begin{alignat*}{4}
&\text{simplicity:}~~~ & \tilde{B}^{\mu \nu}_{[IJ} \tilde{B}^{\mu \nu}_{KL]}&=0& & \mu, \nu & \text{distinct,}\\
&\text{intersection:} & \tilde{B}^{\mu \nu}_{[IJ} \tilde{B}^{\nu \rho}_{KL]}&=0& & \mu, \nu, \rho & \text{distinct,}\\
&\text{normalisation:}~~~ & \tilde{B}^{\mu \nu}_{[IJ} \tilde{B}^{\rho \sigma}_{KL]}&=
\tilde{B}^{\mu \rho}_{[IJ} \tilde{B}^{\sigma\nu}_{KL]} ~~~
& & \mu, \nu, \rho, \sigma & ~ \text{distinct,}
\end{alignat*}
from which only the first two are relevant for us since the last one does not appear in our case where $B^{\mu \nu}_{IJ} \rightarrow B^{t \nu}_{IJ} \rightarrow \pi^{a}_{IJ}$.

We find it convenient for the following considerations to look at $\pi^a_{IJ}$ as a two form $\pi^a_{IJ} dx^I \wedge dx^J$. It can be shown that for a two-form $B_{IJ}$ 
\be  B_{[IJ} B_{KL]} = 0 ~ \Leftrightarrow ~B_{IJ} = u_{[I}v_{J]} \text{,}\ee 
which corresponds to the simplicity constraint. Therefore, all $\pi^{a}_{IJ}$ factor into $u^a_{[I} v^a_{J]}$ (no summation). To complete the proof, we have to relate the different $u^a_I$ to each other. For this purpose, it is proved in \cite{FreidelBFDescriptionOf} that for two two-forms $B_{IJ}$ and $B'_{IJ}$,
\be  B_{[IJ} B'_{KL]}  = 0 ~ \Leftrightarrow ~ B_{IJ} = u_{[I} v_{J]} ~ \text{and} ~ B'_{IJ} = u_{[I} w_{J]} \text{,}\ee 
meaning that the two two-forms share a common factor which is unique up to scaling. In our case, this relation is ensured by the intersection constraint where $\nu=t$.

Combining these two arguments, we realise that $\pi^{a}_{IJ}$ factors into one-forms with a common factor. Introducing the correct density weight and a suitable normalisation, we obtain
\be  \pi^{a}_{IJ} = \pm 2 \sqrt{q} n_{[I} \m^\| e^{a}_{J]} \text{.} \ee
The property of $n^I$ being time-like in the Lorentzian case will be enforced by another constraint. 

The sign can be absorbed into  $n^{I}$ for $D+1$ even, the otherwise appearing signs can be absorbed into the Lagrange multipliers in the Hamiltonian. We remark that in the general case discussed in \cite{FreidelBFDescriptionOf}, the normalisation constraints are necessary and the proof becomes considerably longer. 

Without the normalisation constraints and smeared over space, (\ref{eqn_theoremfreidelcondition}) reduces to
\be  S^{ab}_{\overline{M}} \left[ c_{ab}^{\overline{M}} \right] = \int_\sigma d^Dx \,\frac{1}{4} c_{ab}^{\overline{M}} \epsilon_{IJKL\overline{M}} \pi^{a IJ} \pi^{b KL} \text{.}\ee 
The Lagrange multiplier $c_{ab}^{\overline{M}}$ can be chosen to be symmetric in the index pair $ab$.

\subsection{Constraint Analysis}
\label{sec:ConstraintAnalysis}
We start with the action 
\be  S= \int_\mathbb{R} dt \, \int_\sigma d^Dx \, \left(\frac{1}{2} \pi^{aIJ} \dot{A}_{aIJ}- N \mathcal{H} - N^a \mathcal{H}_a -\frac{1}{2} \lambda_{IJ} G^{IJ} - c_{ab}^{\overline{M}} S^{ab}_{\overline{M}} \right) \text{,} \ee 
where we used the notation from the previous section.  The action is motivated by \cite{FreidelBFDescriptionOf}, but we can also arrive at it by taking the action from the previous chapter, dropping the variables $E^{aI}$, $n_I$, and all constraints containing them, and introducing the BF-simplicity constraint. Since we want the metric to be positive definite, we impose the constraint
\be  s\pi^{aIJ} \pi^{b}_{IJ} \approx 2 q q^{ab} > 0 \text{,} \ee
 where the greater sign means positive definiteness of matrices. In the Lorentzian case, the relation is only satisfied if $n^I$ is time-like, because $E^{aI} E^{b}_I$ would be indefinite otherwise. Such a constraint is called non-holonomic and does not reduce the degrees of freedom.
 
From the above action ``we read off'' the non-vanishing Poisson brackets as $\{A_{aIJ}, \pi^{bKL} \} = 2 \delta_a^b \delta^K_{[I} \delta^L_{J]}$. To see that this is correct, i.e. to read off the 
symplectic structure rather than going through the constraint analysis, consider the following generic situation. All variables appearing in the action have to be considered as configuration variables
and corresponding velocities to begin with. Since 
\be  P_{A_{aIJ}} = \frac{\delta L}{\delta \dot{A}_{aIJ}}=  \pi^{aIJ} \ee 
cannot be solved for $\dot{A}_{aIJ}$, $P_{A_{aIJ}} - \pi^{aIJ} \approx 0$ becomes a constraint and has to be added to the Hamiltonian and smeared with a Lagrange multiplier $\mu_{aIJ}$. The same is true for
\be  P_{\pi^{aIJ}} = \frac{\delta L}{\delta \dot{\pi}^{aIJ}}= 0 \text{.} \ee 
We will call the corresponding multiplier $\nu^{aIJ}$. Before proceeding with the canonical analysis, we can use $P_{A_{aIJ}} - \pi^{aIJ} \approx 0$ to substitute all $\pi^{aIJ}$ for $P_{A_{aIJ}}$ by redefining $\mu_{aIJ}$. Stability of $P_{A_{aIJ}} - \pi^{aIJ} \approx 0$ requires to adjust $\nu^{aIJ}$. Stability of $P_{\pi^{aIJ}} \approx 0$ requires us to set $\mu_{aIJ} = 0$. The stability of the other constraints is equivalent to the case where we ``read off'' the symplectic structure and we assume it to be satisfied. We can now solve the constraints $P_{A_{aIJ}} - \pi^{aIJ} \approx 0$ and $P_{\pi^{aIJ}} \approx 0$ and perform a symplectic reduction which gives us the new Poisson bracket 
\be  \{A_{aIJ}, \pi^{bKL} \} =  2 \delta_a^b \delta^K_{[I} \delta^L_{J]} \text{.} \ee 

Most of the canonical analysis is the same as in the previous chapter and we will only describe the differences. The Poisson bracket 
\begin{eqnarray} 
\left\{ \mathcal{H}[M],  \mathcal{H}[N] \right\}  & = &   -\frac{1}{2} \mathcal{H}_a \left[(M\partial_bN-N\partial_bM) \pi^{aIJ}\pi^b_{IJ} \right] \nonumber \\ 
& &+s\frac{1}{4} S^{ab}_{\overline{M}} \left[(M\partial_aN-N\partial_aM) \epsilon_{IJKL} \m^{\overline{M}}\pi^{cIJ}F_{cb}^{KL} \right] 
\end{eqnarray}
of two Hamiltonian constraints reproduces exactly the BF-simplicity constraint and shows that the theory would be inconsistent without this constraint. The BF-simplicity constraint is stable under spatial diffeomorphisms and internal rotations as reflected by the Poisson brackets
\be  \left\{ \tilde{\mathcal{H}}_a[N^a],  S^{ab}_{\overline{M}}[c_{ab}^{\overline{M}}] \right\}  =  - S^{ab}_{\overline{M}} \left[ (\mathcal{L}_N c)_{ab}^{\overline{M}} \right] \ee 
and 
\be  \left\{ \frac{1}{2} G^{IJ}[\lambda_{IJ}],  S^{ab}_{\overline{M}}[c_{ab}^{\overline{M}}] \right\}  =  S^{ab}_{\overline{M}}\left[\sum_{i=1}^{D-3} \lambda^{M_i} \m_{M'_i} c_{ab}^{M_1 \ldots M_{i-1} M'_i M_{i+1} \ldots M_{D-3}}\right] \text{,}\ee 
and trivially commutes with itself. As in the previous chapter, the Poisson bracket with the Hamiltonian constraint 
\be  \left\{ S^{ab}_{\overline{M}}[c_{ab}^{\overline{M}}], \mathcal{H}[N] \right\}  =  D^{ab}_{\overline{M}} \left[Nc_{ab}^{\overline{M}} \right] + S^{ab}_{\overline{M}}[\ldots] \ee 
imposes a new constraint 
\be D^{ab}_{\overline{M}} = -\epsilon_{IJKL\overline{M}} \pi^{cIJ} \left( \pi^{(a|KN} D_c \pi^{b)L}\m_N \right) \text{.} \ee 
To show its stability, we have to show that its Poisson bracket with the BF-simplicity is invertible. 
Irrespective of this, we emphasise that $D^{ab}_{\overline{M}}$ is stable under internal rotations, reflected by 
\be  \left\{ \frac{1}{2} G^{IJ}[\lambda_{IJ}],  D^{ab}_{\overline{M}}[d_{ab}^{\overline{M}}] \right\}  =  D^{ab}_{\overline{M}}\left[\sum_{i=1}^{D-3} \lambda^{M_i} \m_{M'_i} d_{ab}^{M_1 \ldots M_{i-1} M'_i M_{i+1} \ldots M_{D-3}}\right] \text{.}\ee 
Concerning the diffeomorphism constraint, it is easy to see that we can extend the covariant derivative in $D^{ab}_{\overline{M}}$ to act on spatial indices via the Christoffel symbols. 
Namely, adding the corresponding terms to the constraint, we see that, due to the symmetry of the Christoffel symbols in their lower indices, the added terms are proportional to simplicity constraints. $D^{ab}_{\overline{M}}$ therefore transforms like a scalar density of weight $+3$ under spatial diffeomorphisms and the Poisson bracket with the diffeomorphism constraint has to be proportional to the $D^{ab}_{\overline{M}}$. Another easy way to do this calculation is to use the Jacobi identity after expressing $D^{ab}_{\overline{M}}$ as a Poisson bracket. We do not know of any nice way to express the Poisson bracket of $D^{ab}_{\overline{M}}$ with the Hamiltonian constraint and will leave the discussion of this bracket open, as its value is not important in the following. 

Counting of the degrees of freedom which the BF-simplicity constraint reduces, i.e. $\pi^{aIJ} \rightarrow E^{aI}$, yields \mbox{$D^2(D-1)/2 - D$} which is for $D>3$ less than the number of Lagrange multipliers \mbox{ $\frac{1}{2}D(D+1)\binom{D+1}{4}$}. The BF-simplicity constraints are therefore not independent and the matrix formed by calculating the Poisson bracket with $D^{ab}_{\overline{M}}$ cannot be invertible. The solution to this problem is to find an independent set of BF-simplicity and $D$ constraints which still enforce the same constraint surface. The constraints of the previous chapter do have this property and lead us to the following Ansatz. We choose some internal time-like vector $n^I$ with $n^In_I = s$ which may vary as a function of the spatial coordinates and decompose $\pi^{aIJ}$ as
\be  \pi^{aIJ} = \bar{\pi}^{aIJ} + 2 n^{[I}E^{a|J]} \ee 
as in the previous chapter. We also define $E_{aI}$ by $E_{aI} E^{bI} = \delta^b_a$ and $n^IE_{aI}=0$. Together with the normalisation condition $n^I n_I=s$ this means that $n^I=n^I[E]$ 
can be considered as a function of $E^a_I$ only and thus does not count as independent 
degree of freedom. The BF-simplicity constraints plus the non-holonomic constraint are equivalent with $\bar{\pi}^{aIJ} = 0$ and $n_I$ being time-like in the Lorentzian case. However, $\bar{\pi}^{aIJ}$ has $D^2(D-1)/2$ degrees of freedom and $E^a_I$
has $D(D+1)$ which together yields $D^2(D+1)/2+D$ degrees of freedom while $\pi^{aIJ}$ 
has only $D^2(D+1)/2$ degrees of freedom. It follows that $\bar{\pi}^{aIJ}$ and $E^a_I$ cannot 
be considered as independent degrees of freedom, there must be $D$ additional relations
among them. Indeed, in the companion paper \cite{BTTI} we will argue\footnote{This is not trivial:
For $D\ge 3$ one cannot use closed formulas for a proof. It is apparently necessary to make use of fixed point theorems.}  that 
it is always possible to arrange that $\bar{\pi}^{aIJ}=\bar{\pi}^{aIJ}_T$ is automatically 
trace free with respect to $E_{aI}$. These would be the missing $D$ relations and now the BF-Simplicity constraints are equivalent with the $D^2(D-1)/2-D$ constraints $\bar{\pi}^{aIJ}_T=0$
which in number match with the constraints $\bar{K}^T_{aIJ}=0$ to which the constraints
$D_{\overline{M}}^{ab}=0$ reduce as we will now show below. 
Presently not having a proof for $D\ge 3$ that such a decomposition
is possible on the full phase space, we restrict to the part
of the phase space where this is the case.
Our approach will be legitimated at the end of this section. In other words, we will only allow $\pi^{aIJ}$ of the following form:
There is a tensor $E^{aI}$ with $Q^{ab}=\eta_{IJ} E^{aI} E^{bJ}$ positive definite. Let $n^I[E]$ be 
the unique normal vector satisfying $E^{aI} n_I=0,\; n^I n_I=s$. Take any tensor $t^{aIJ}$ 
and construct from it $\bar{t}^{aIJ}_T[t,E]$ using $E,n[E]$. Then $\pi^{aIJ}:=\bar{t}^{aIJ}_T+2 n^{[I} E^{aJ]}$
and automatically $E^{aI}= -s \pi^{aIJ} n_J$. For $\pi^{aIJ}$ constructed in this way, the fixed 
point equation derived in \cite{BTTI} has obviously non trivial solutions and 
the question is whether such $\pi^{aIJ}$ are generic.     

Concerning the $D_{ab}^{\overline{M}}$ constraint, we make the same Ansatz as in the previous chapter and set
\be  A_{aIJ} = \Gamma_{aIJ} + \bar{K}_{aIJ} + 2 n_{[I}K_{a|J]} \text{.} \ee 
A short calculation yields
\begin{eqnarray} \label{multiplier}
& & 
\bar{f}_{(a|IJ}\pi_{|b)KL} \epsilon^{IJKL \overline{M}} D_{\overline{M}}^{ab} 
=-\bar{f}_{(a|IJ}\pi_{|b)KL} \epsilon^{IJKL \overline{M}} \epsilon_{ABCD\overline{M}} \pi^{cAB} \pi^{(a|C} \m_E D_c \pi^{b)DE}
\nonumber \\
 & \approx & -(D-3)!(D-1) \bar{K}_{aIJ}   F^{aIJ,bKL} \bar{f}_{bKL} \text{,}
\end{eqnarray}
where $F^{aIJ,bKL}$ denotes the same matrix as in (\ref{tensor}). We defined $\pi_{bKL} := q^{-1} q_{ab} \pi^{a}\m_{KL}$, where $q^{-1} q_{ab}$ is the inverse matrix of $\frac{s}{2} \pi^{aIJ} \pi^{b}\m_{IJ}$, such that $\pi^{aIJ} \pi_{bIJ} = 2s \delta^a_b$. We notice that $\bar{f}_{aIJ}$ can be chosen traceless with respect to $E^{aI}$, since any trace part would drop out in the combination $\bar{f}_{(a|IJ} \pi_{b)KL} \epsilon^{IJKL \overline{M}}$ modulo the BF-Simplicity constraint. The subset of $D$ constraints parametrised by $\bar{f}_{aIJ}=\bar{f}^T_{aIJ}$ as above thus sets the trace free part of $\bar{K}_{aIJ}$ to zero. When inserting the solution of the BF-simplicity constraint into the full $D_{\overline{M}}^{ab}$ constraint, we get
\be  D_{\overline{M}}^{ab} \approx - 2 s \epsilon_{ABC} \m^D \m_{\overline{M}} n^AE^{cB}E^{(a|C}E^{b)E} \bar{K}_{cED} \ee 
and immediately verify that the solution $\bar{K}^T_{aIJ}=0$ solves all the $D$ constraints
because the trace part of $\bar{K}_{cED}$ drops out in the above combination. 

From these considerations, we realise that it is legitimate to use the Lagrange multipliers 
displayed in (\ref{multiplier}) and therefore only a subset of the $D$ constraints. It follows that we only have to check the stability of this subset of constraints. To form the Dirac matrix, we choose 
similarly a subset of BF-simplicity constraints equivalent to $\bar{\pi}^{aIJ}_T=0$ and calculate
\begin{eqnarray}
& & 
\int\; d^Dx\; \int\; d^Dy\;
[\bar{f}^T_{(a|IJ}\pi_{b)KL}\epsilon^{IJKL\overline{M}}](x) \left\{ S^{ab}_{\overline{M}}(x), D^{cd}_{\overline{N}}(y) \right\}  [\bar{g}^T_{(c|MN}\pi_{d)OP}\epsilon^{MNOP\overline{N}}](y) 
\nonumber \\
& \approx & 4 (D-1)^2 ((D-3)!)^2\int\; d^Dx\;   \bar{f}^T_{aIJ} F^{aIJ,bKL} \bar{g}^T_{bKL}  \text{.}
\end{eqnarray}
We can therefore adjust the multiplier of the BF-simplicity such that the independent subset of $D$ constraints is stable under time evolution and finish the canonical analysis. 
Since the Dirac matrix is invertible, the chosen subset of BF-simplicity constraints has to be independent. The number of BF-simplicities in this subset is equivalent to the number of degrees of freedom in a transverse trace free matrix, i.e. $D^2(D-1)/2-D=D(D+1)(D-2)/2$ and matches the degrees of freedom which are to be taken out of the system by the full BF-simplicity constraints and all BF-simplicity constraints can thus be derived by taking the linear span of this subset. 

The solution of the constraints proceeds analogously to the previous chapter, the only difference being that we do not need to solve the momenta associated with $E^{aI}$ and $n^I$. The two formulations presented are therefore equivalent.

\subsection{Degrees of Freedom}

As in the previous chapter, we check the degrees of freedom of the Hamiltonian system derived using the BF-simplicity constraint. For $\mathcal{H}$ to become a first class constraint, we construct the linear combination (using the same abuse of notation as before)
\be \tilde{\mathcal{H}} := \mathcal{H} 
- \int \;\bar{S}^{aIJ}_T \left( \left\{ \bar{D}_T, \bar{S}_T \right\}^{-1} \right)_{aIJ, bKL} \left\{\bar{D}^{bKL}_T, \mathcal{H} \right\} \text{.} \ee   
Since the Dirac matrix between the independent BF-simplicity and $D$ constraints is invertible, they are of the second class. The rest of the constraints is of the first class. 
\mbox{} \\
\begin{center}
\begin{tabular}[c]{ c | c || c | c  }
  Variable & DoF &  Constraint & DoF \\
 \hline
\hline
  $A_{aIJ}$ & $ \frac{D^2(D+1)}{2}$ & First class & (count twice!)\\
  $\pi^{aIJ}$ & $ \frac{D^2(D+1)}{2}$ & $\tilde{\mathcal{H}}$ & $1$ \\
  $\lambda_{IJ}$ & $\frac{D(D+1)}{2}$ & $\mathcal{H}_a$  &$ D$ \\
  $N$ & $1$ & $G^{IJ}$ & $\frac{D(D+1)}{2}$ \\
  $N^a$ & $D$ & $P_N$  & $1$ \\
  $P_{\lambda_{IJ}}$ & $\frac{D(D+1)}{2}$ & $P_{N^a}$ &$D$ \\
  $P_N$ & $1$ & $P_{\lambda_{IJ}}$  &  $\frac{D(D+1)}{2}$  \\  \cline {3-2} \cline{4-2}
  $P_{N^a}$ & $D$ & Second class & \\
   &  &  $S^{ab}_{\overline{M}}$ & $\frac{D^2(D-1)}{2}-D$ \\
   &  & $D^{ab}_{\overline{M}}$ & $\frac{D^2 (D-1)}{2}-D$ \\
  \hline \hline
  Sum: & $D^3+2D^2+3D+2$ & Sum: & $D^3 + D^2 + 4D+4$ 
\end{tabular} 
\end{center}
\mbox{} \\
The difference between the degrees of freedom and the weighted sum of the constraints is again $(D+1)(D-2)$ and matches those of General Relativity.

\section{Review of Gauge Unfixing}
\label{section_gu}

The name ``gauge unfixing'' suggests that this is a procedure in some sense 
inverse to ``gauge fixing''. To see to what extent this is indeed the case it 
is useful to recall some facts about gauge fixing first. After that we 
focus on the gauge unfixing case. This review section 
can be skipped by readers familiar with gauge (un)fixing although we add a few 
extra twists to it.
We have combined material from several sources:
To the best of our knowledge, the pioneering paper on gauge unfixing of second class theories 
is \cite{MitraGaugeInvariantReformulationAnomalous} and the 
general theory was developed in \cite{AnishettyGaugeInvarianceIn, VytheeswaranGaugeUnfixingIn}. 
Parts of this theory were independently rediscovered from the point of view 
of a first class theory in \cite{HenneauxQuantizationOfGauge, DittrichPartialAndComplete}, see 
also \cite{RovelliWhatIsObservable, RovelliPartialObservables, ThiemannReducedPhaseSpace}.

\subsection{Gauge Fixing}

Recall that gauge fixing of a 
{\it first class system} with first class constraints $S_I$ (where $I$ 
takes values in some index set) on a phase space $\cal M$ consists in imposing
an equal number of gauge fixing conditions $D_I$ such that the matrix $M$ with 
entries $M_{IJ}:=\{S_I,D_J\}$ is regular. The gauge fixing conditions, modulo
the problem of Gribov copies, select a unique point on each gauge orbit 
of the $S_I$. Here the gauge orbit of a point $m\in {\cal M}$ is the 
set\footnote{In case that the first class constraints close with non trivial 
structure functions only, it maybe necessary to apply several of the Poisson 
automorphisms $\alpha_\beta$ with different $\beta$ because the $\alpha_\beta$
do not form a group under concatenation in this case.}
\be \label{1}
[m]:=\{\alpha_\beta(m),\;\beta^I\in \mathbb{R}\},\;\;\;\;
\alpha_\beta(f):=\exp(\beta^I\{S_I,.\})\cdot f \text{,}
\ee
where $\alpha_\beta(f)$ is the gauge flow with parameter $\beta$ applied to the 
(smooth) function $f$ on phase space. To qualify as an admissible gauge fixing
condition, at least on the constraint surface 
\be \label{2}
\overline{{\cal M}}:=\{m\in {\cal M};\;S_I(m)=0\;\forall\; I\} \text{,}
\ee
it must be possible to reach the selected section 
\be \label{3}
\sigma_D(\overline{{\cal M}}):=\{m\in \overline{{\cal M}};\;
D_I(m)=0\;\forall\; I\}
\ee
from any other section of $\overline{{\cal M}}$. 

At least locally, the 
constraint surface acquires the structure of a fibre bundle where the fibres
are given by the gauge orbits (considered as subsets of 
$\overline{{\cal M}}$) and the base space is the set of equivalence classes 
\be \label{4}
\widehat{{\cal M}}:=\{[m];\;\;m\in \overline{{M}}\}
\ee
called the reduced phase space.
Under the above conditions there is a bijection between 
$\sigma_D(\overline{{\cal M}})$ and $\widehat{M}$: Given 
$m\in \sigma_D(\overline{{\cal M}})$ one obtains $[m]\in \widehat{{\cal M}}$
via (\ref{1}) and given $[m]$ (considered as a subset of 
$\overline{{\cal M}}$) one computes the unique point $m'\in [m]$ such that
$D_I(m')=0$ for all $I$, that is $m'=[m]\cap \sigma_D(\overline{{\cal M}})$. 
However, while the construction of $\widehat{{\cal M}}$ 
is {\it canonical}, i.e. does not use any structure other than $S_I$ which 
canonically follow from the Dirac algorithm applied to the singular 
Lagrangian in question, the cross section 
$\sigma_D(\overline{{\cal M}})$ uses the additional input of $D$ which,
except for the regularity condition on $M$, is rather arbitrary. 

The observables of the first class system are the gauge invariant functions
evaluated on the constraint surface. These therefore only depend on the 
equivalence classes $[m]$. It appears that the construction of such gauge 
invariant functions is generically impossible for sufficiently complicated 
constraints $S_I$. This turns out to be correct if one is interested in 
these observables as functions on $\overline{\cal M}$. However, given a set
of gauge fixing conditions $D_I$, not only can one write an explicit formula
for these observables but one can also compute their Poisson algebra. 
This also then displays the relation between the spaces 
$\sigma_D(\overline{{\cal M}})$ and $\overline{{\cal M}}$ in explicit form.
Given a function $f$ on $\cal M$ one can define a weak Dirac observable 
by the formula
\be \label{5}
O^{(D)}(f):=[\alpha_\beta(f)]_{\alpha_\beta(D)=0} \text{,}
\ee
where the superscript $(D)$ is to make it explicit that this formula is 
not canonical but depends on the chosen gauge fixing.
This formula has to be understood in the following way: First one computes 
the gauge flow of $f$ at $m\in \overline{{\cal M}}$ 
with real valued (phase space independent) constants
$\beta^I$, that is
\be \label{6}
\alpha_\beta(f):=f+\sum_{n=1}^\infty\; \frac{1}{n!}\; \beta^{I_1}..\beta^{I_n}
\; \{S_{I_1},\{..,\{S_{I_n},f\}..\}\}
\ee
and then one solves the condition $\alpha_\beta(D_I)=0$ for all $I$ for 
$\beta^I=\gamma^I(m)$ and inserts the corresponding phase space dependent 
function into (\ref{6}). The value $\gamma(m)$ is thus the parameter needed 
in order to map $m$ to that point on its orbit $[m]$ at which the $D_I$ 
vanish. It is not difficult to check that indeed $\{S_I,O_f\}\approx 0$,
and that $O^{(D)}$ 
preserves the pointwise addition and multiplication of functions
\be \label{7}
O^{(D)}({f+g})=O^{(D)}(f)+O^{(D)}(g),\;\;\;\; O^{(D)}({fg})=O^{(D)}(f)\; O^{(D)}(g) \text{.}
\ee
Moreover, the following remarkable formula holds\footnote{The first 
identity holds because the $S$ constraints form a subalgebra. The Dirac
matrix $M_{\alpha\beta}=\{C_\alpha,C_\beta\},\;\{C_\alpha\}=\{S_I,D_I\}$
on the constraint surface therefore has the symbolic structure
$M=\left( \begin{array}{cc} 0 & B\\ -B & A \end{array} \right)$
and its inverse is given by 
$M^{-1}=\left( \begin{array}{cc} B^{-1} A B^{-1} & -B^{-1}\\ 
B^{-1} & 0 \end{array} \right)$ so that the Dirac bracket $\{F,G\}^\ast$
contains no terms $\propto \{F,D\}\{G,D\}$.}
\be \label{8}
\{O^{(D)}(f),O^{(D)}(g)\}\approx\{O^{(D)}(f),O^{(D)}(g)\}^\ast_{S,D}
\approx O^{(D)}({\{f,g\}^\ast_{S,D}}) \text{,}
\ee
where $\{.,.\}^\ast_{S,D}$ is the Dirac bracket of the {\it second class system}
of constraints $S_I,D_I$. Since a sufficient number of the $O^{(D)}(f)$
serves as coordinates of $\widehat{{\cal M}}$ we see that the Poisson bracket
on the reduced phase space $\widehat{{\cal M}}$ is given by the Dirac bracket
and $O^{(D)}$ is a Dirac bracket homomorphism from the algebra of 
smooth functions on $\cal M$ to the one on $\widehat{{\cal M}}$. 

It should 
be stressed, however, that this algebra of observables is not canonical,
it depends on the choice of admissible gauge fixing $D$ which is an extra 
input necessary for their very construction. Nevertheless, once we have made 
such a choice, we see that a first
class system $S$ together with a gauge fixing condition $D$ 
is completely equivalent to the second class system $S,D$. Namely, for a
second class system the reduced phase space consists simply in the constraint
manifold 
\be \label{9}
\overline{\overline{{\cal M}}}:=\{m\in {\cal M};\;\;S_I(m)=D_I(m)=0\;\forall 
\; I\}\equiv \sigma_D(\overline{{\cal M}}) \text{,}
\ee
which precisely coincides with the gauge section (\ref{3}), and the 
symplectic structure on $\overline{\overline{{\cal M}}}$ is given by the 
Dirac bracket
\be \label{10}
\{f,g\}^\ast_{S,D}=\{f,g\}-\{f,C_\alpha\} \; 
[M^{-1}]^{\alpha\beta}\; \{C_\beta,g\} \text{,}
\ee
where $\{C_\alpha\}=\{S_I,D_I\}$ and $M_{\alpha\beta}=\{C_\alpha,C_\beta\}$
is non degenerate by construction. When restricting $O^{(D)}$ to 
$\overline{\overline{{\cal M}}}$ which is in bijection with 
$\widehat{{\cal M}}$ as we have seen, it becomes a Dirac bracket isomorphism.

\subsection{Gauge Unfixing}

We now consider a second class system with constraints $S_I,D_I$ with the 
special structure that $S_I$ is a first class subalgebra of constraints,
that is 
\be \label{11}
\{S_I,S_J\}=f_{IJ}\;^K\;S_K
\ee
for certain structure functions $f_{IJ}\;^K$ and $M_{IJ}:=\{S_I,D_J\}$
is supposed to be non degenerate on the constraint surface 
\be \label{12}
\overline{\overline{{\cal M}}}:=\{m\in {\cal M};\;\;S_I(m)=D_I(m)=0\;\forall 
\; I\} \text{,}
\ee
which is equipped with the Dirac bracket (\ref{10}). In 
\cite{VytheeswaranGaugeUnfixingIn} we 
find conditions under which linear combinations of a given set of second 
class constraints can be subdivided into sets $S_I$ and $D_I$ subject
to (\ref{11}). Here we will simply assume that this has been achieved.

We have seen in the previous section that a first class system $S_I$ together
with additional gauge fixing conditions $D_I$ is equivalent with the second 
class system $S_I,D_I$. The idea of gauge unfixing is now simply to interpret
the given second class system of constraints $S_I,D_I$ as just a first class 
system $S_I$ to which the particular gauge fixing conditions $D_I$ have 
been added. 

This point of view has the following advantage towards quantisation:\\ 
For a first 
class system of constraints we have two possible quantisation strategies,
namely \mbox{A. Operator} Constraint Quantisation and B. Reduced Phase Space 
Quantisation. The advantage of option A. is that one can work with the 
simple Poisson bracket algebra on the kinematical phase space $\cal M$ for 
which Hilbert space representations are typically easy to find and the task
is to find those which support the $S_I$ as densely defined, closable and 
non anomalous operators. The disadvantage is that one has to equip the 
joint kernel of the constraints with a new (physical) inner product which 
carries a representation of the observables of the theory and while there 
are general tools available such as group averaging, it is generically not
possible to determine the physical Hilbert space in closed form.  
The disadvantage of option B. is that the Dirac bracket algebra 
on the reduced phase space is typically so complicated that no Hilbert space
representations can be found. On the other hand, if one manages to do so,
then one has direct access to the physical Hilbert space and the algebra 
of observables. Now in case that option B. is inhibited due to the complexity
of the Dirac bracket algebra which is typically the case for second 
class systems, option A. appears to be the only possible approach to 
quantisation. As we will see, one can do even better than that, but let us 
assume for the moment that we take a second class system $S_I,D_I$ 
with complicated Dirac bracket algebra and therefore drop $D_I$ and 
just perform an operator constraint quantisation of $S_I$.     
 
At first sight, this strategy seems to be false for at least two reasons:
\begin{itemize}
\item[A)] 
From the point of view of the first class system, the gauge fixing 
conditions $D_I$ are just one of an infinite number of possible choices,
the first class system does not know about the $D_I$ and therefore one 
can drop the $D_I$. However, we are not given a first class system, we 
are given a second class system and 
from the point of view of the second class system, the $D_I$ are 
{\it canonical},
they follow canonically from Dirac's stabiliser algorithm applied to the 
given singular Lagrangian. It seems therefore to be wrong to forget about 
the special role of the $D_I$ within the first class system as we would drop
information that is forced on us by Dirac's algorithm. However, imposing
the $D_I$ as operators as well in the quantum theory is not possible, that
is, the joint kernel of the $D_I,S_I$ is just the zero vector.
\item[B)]
The canonical Hamiltonian $H$ of the second class system as derived via Dirac's
stabiliser algorithm is typically not gauge invariant with respect to the 
$S_I$ which would not be the case for a true first class system with just
the constraints $S_I$. In fact, in many applications the second class 
structure $S_I,D_I$ arises from primary constraints $S_I$ and a canonical
Hamiltonian of the form 
\be \label{13}
H'=H_0+\lambda^I \; S_I \text{,}
\ee
with nontrivial $H_0$ independent of the $S_I$ (that is $[H_0]_{S=0}\not=0$) 
and the $D_I$ arise as secondary constraints from the stability requirement
\be \label{14}
0\stackrel{!}{=}\{H',S_I\}\;\approx\;\{H_0,S_I\}=:D_I \text{,}
\ee
where $\{S_I,S_J\}\propto S_K\approx 0$ was used. 
The stability of the $D_I$ fixes the 
Lagrange multipliers $\lambda^I$
\be \label{15}
0\stackrel{!}{=}\{H',D_I\}=\{H_0,S_I\}+\lambda^J\{S_J,D_I\}\;\;
\Rightarrow\;\;\lambda^I=-[M^{-1}]^{JI}\;\{H_0,S_J\}=:\lambda_0^I \text{,}
\ee
so that the stabilised, first class Hamiltonian (it weakly commutes
with all the constraints $S_I,D_I$) reads
\be \label{16}
H=H_0+\lambda_0^I \; S_I \text{.}
\ee
It is not gauge invariant with respect to just the constraints $S_I$ since 
$\{H,S_I\}\approx D_I$ so that the constraints $D_I$ appear again as 
a consistency condition.
\end{itemize}
We now explain how both obstacles can be overcome. We deal first with the 
second issue B): We simply make the canonical Hamiltonian $H$ gauge invariant
with respect to the $S_I$ by using the map $O^{(D)}$ displayed in (\ref{5})
that is, we replace $H$ by 
\be \label{17}
\tilde{H}:=O^{(D)}(H) \text{.}
\ee
To see that this is an allowed Hamiltonian within the second class system 
we need to compute $\tilde{H}$ in some detail. As one can show 
\cite{DittrichPartialAndComplete, ThiemannReducedPhaseSpace} one has explicitly
\be \label{18}
O^{(D)}(H)=H+\sum_{n=1}^\infty\; \frac{1}{n!}\;\prod_{k=1}^n\;[-D_{I_k}]\;
\{S^{\prime I_1},..\{S^{\prime I_n}, H\}..\} \text{,}
\ee
where $S^{\prime I}=[M^{-1}]^{IJ} S_J$ so that 
$\{S^{\prime I},D_J\}=\delta^I_J$ modulo $S$. We have 
\begin{eqnarray} \label{19}
\tilde{H}-H &=&
-D_I\{S^{\prime I},H\}+\mathcal{O}(D^2)=
-D_I \left([M^{-1}]^{IJ}\{S_J,H\}+ \{M^{IJ}, H\} S_J\right)+\mathcal{O}(D^2)
\nonumber\\
&=&
-D_I \left([M^{-1}]^{IJ}[D_J+N_J^K S_K]+ \{M^{IJ}, H\} S_J \right)+\mathcal{O}(D^2)=
\mathcal{O}(D^2,D S)
\end{eqnarray}
for some $N_J^K$.
Therefore $\tilde{H}$ and $H$ differ by terms at least quadratic in the 
constraints and thus do not spoil the first class structure of $H$.
Therefore $\tilde{H}$ is an admissible Hamiltonian for the second class system
which is simultaneously weakly invariant with respect to the $S_I$. This is 
also the reason why one did not choose $\tilde{H}'=O^{(G)}(H)$ for some 
gauge fixing conditions $G_I\not=D_I$ because by a similar calculation 
as in (\ref{19}) one would now compute $\tilde{H}'-H=O(D G,S G,G^2)$
but $G_I$ is no constraint and thus $\tilde{H}'$ is not an admissible 
Hamiltonian for the second class system.  Notice also that $H$ and
$\tilde{H}$ generate the same equations of motion on
$\overline{\overline{{\cal M}}}$.

We now come to issue A). The question is: How can it be that the 
first class constrained Hamiltonian system $(\tilde{H},S_I)$ be equivalent to 
the second class system $(\tilde{H},S_I,D_I)$? The first class system does 
not know about the $D_I$. It is true that if we choose the special gauge 
fixing conditions $G_I:=D_I=0$ for the first class system, then the reduced 
phase spaces of the two systems are indeed isomorphic as we have shown 
above. However, the choice of $G_I$ is arbitrary from the point of view 
of the first class system as long as the matrix with entries $\{S_I,G_J\}$
is non degenerate and therefore it appears that one has to still somehow 
feed the additional
information about the special role of the gauge fixing condition $G_I=D_I$
into the first class system. However, this is not the case:\\
The point is simply that an arbitrary gauge condition $G_I=0$ is related by 
a gauge transformation generated by the $S_I$ to the gauge condition
$D_I=0$. Therefore the observables of the form $O^{(G)}(f)$ are in fact linear 
combinations, with phase space independent coefficients, of the observables 
of the form $O^{(D)}(f)$. This follows
simply from the fact that for gauge invariant functions $F$ (with respect 
to the $S_I$) we have $F\approx O^{(D)}(F)$. Applied to $F=O^{(G)}(f)$
it follows 
\be \label{20}
O^{(G)}(f)\approx O^{(D)} \left(O^{(G)}(f)\right) \text{.}
\ee
Hence any observable of the form $O^{(G)}(f)$ can be written as $O^{(D)}(f')$
for some other function $f'=O^{(G)}(f)$. Since the roles of $G_I,D_I$ can be 
interchanged we see that the range of the maps $O^{(D)},O^{(G)}$ is the same.
Since the algebra of the $O^{(G)}(f)$ and of the $O^{(D)}(f)$ 
can be computed using the original Poisson bracket on the unreduced phase 
space we see that the algebra of the $O^{(D)}(f)$ and $O^{(G)}(f)$ are 
isomorphic, i.e. it does not matter whether we display one and the 
same observable $F$ in the form $F=O^{(D)}(f)$ or in the form 
$F=O^{(G)}(f')$.

What is different are of course the maps $O^{(D)},\; O^{(G)}$
which provide different gauge invariant extensions of a given function   
$f$. Only the map $O^{(D)}$ yields an isomorphism with the Dirac bracket
algebra of the second class system. However, this does not mean that
one cannot use $O^{(G)}$ to construct gauge invariant observables. It just
means that the identification between the Dirac bracket algebra of functions
on $\overline{\overline{{\cal M}}}$ with the Poisson bracket algebra on
$\widehat{{\cal M}}$ is rather complicated to write down because the correct 
gauge invariant function is $O^{(D)}(f)\approx O^{(G)}(O^{(D)}(f))$ and not 
just $O^{(G)}(f)$.\\
\\
Remarks:\\
1.\\     
This last observation now is also the key to a reduced phase space quantisation 
approach to second class systems $(H,S_I,D_I)$: After having replaced 
it by an equivalent first class system $(\tilde{H},S_I)$ one can now 
make use of the local Abelianisation theorem (see e.g. \cite{GieselAQG4} and 
references therein) and replace the 
constraints $S_I$ by an equivalent, strongly Abelian set 
$S'_I=\pi_I+h_I(\phi^I;q^a,p_a)$ at least locally in phase space 
where the system of first class 
constraints $S_I$ has been solved for some of the momenta $\pi_I$ in terms 
of its conjugate variables $\phi^I$ and the remaining canonical pairs 
$(q^a,p_a)$. Using the natural gauge fixing condition $G_I=\phi_I$ 
the algebra of the $Q^a:=O^{(G)}(q^a),\;P_a:=O^{(G)}(p_a)$ coincides with 
the algebra of the $q^a,p_a$ because the corresponding Dirac bracket does not
affect the subalgebra of functions of $q^a,p_a$. Since the 
algebra of the $Q^a,P_a$ is simple it can be quantised. This is 
surprising because we could have chosen to solve the constraints $S_I=D_I=0$
for $S'_I=\pi_I-\Pi_I(q^a,p_a),\;D'_I=\phi^I-\Phi^I(q^a,p_a)$ from the 
outset so 
that the reduced phase space is parametrised by the $q^a,p_a$ but 
the corresponding Dirac bracket $\{p_a,q^b\}^\ast\not=\delta_a^b$
is not simple. The reason is of course that the functions 
$Q^a,P_a$ are genuinely different from $q^a,p_a$, in fact they are nontrivial
functions of $\phi^I,q^a,p_a$ built in such a way that they have 
a simple Dirac bracket with respect to $S,D$. Moreover 
$\{Q^a,P_b\}=\{Q^a,P_b\}^\ast_{S,D}$ due to gauge invariance.
This holds for any two pairs of gauge invariant functions, in particular
for the Hamiltonian $\tilde{H}$. \\
2.\\
For generally covariant systems $H_0$ is not a true Hamiltonian but rather
a linear combination of different constraints $H_0=\mu^A C'_A$, typically
a closed subalgebra of the form $\{C'_A,C'_B\}=f_{AB}\;^C \; C'_C$ such that 
$\{C'_A,S_I\}=f_{AI}\;^J S_J$ for $A\not=0$ and $\{C'_0,S_I\}=D_I$ thus 
$\{H_0,S_I\}\approx \mu^0 D_I$. 
In our applications it will turn out that 
$\{C'_A,D_I\}=\tilde{f}_{AI}\;^K D_K,\; A\not=0$ and $\{C'_0,D_I\}$ is not 
weakly zero.
The Dirac stabiliser algorithm then 
replaces $C'_0$ by $C_0=C'_0-M^{JI}\{C'_0,D_J\} S_I$ so that 
$\{C_0,D_I\}=0$ while $C_A'=C_A$ 
for $A\not=0$ and $H'$ is replaced by $H=\mu^A C_A$. The $C_A$ now  
close among themselves modulo $S_I$. Application of $O^{(D)}$ replaces 
$H$ by $\tilde{H}=\mu^A\tilde{C}_A,\;\tilde{C}_A=O^{(D)}(C_A)$. 
Now
modulo $S_I$ constraints 
\be \label{21}
\{\tilde{C}_A,\tilde{C}_B\}\approx O^{(D)}(\{C_A,C_B\}^\ast_{S,D})
\ee
and 
\begin{eqnarray} \label{22}
\{C_A,C_B\}^\ast_{S,D} &\propto& \{C_A, C_B\},\;\;\;
\{C_A,S_I\}\{C_B, S_J\},\;   \;\;
\{C_A,S_I\}\{C_B, D_J\},\;\;\;
\{C_A,D_I\}\{C_B, D_J\},\;
\nonumber\\
&\propto& C_A, S_I, D_I \text{.}
\end{eqnarray}
Since $O^{(D)}(D_I)\approx 0$ it follows that the $\tilde{C}_A$ and the 
$S_I$ form a first class algebra.\\    
3.\\
Whether gauge unfixing is feasible depends largely on the question whether
the series that determines $\tilde{H}$ terminates. Fortunately, in our 
application this will be the case. \\
4.\\
The formula $O^{(G)}(f)\approx O^{(D)}(O^{(G)}(f))$ does not display the fact 
that $G$ can be reached from $D$ via a gauge transformation. However, using
the fact that $O^{(D)}(D_I)\approx 0$ and that $O^{(G)}(f)$ is a power series 
in $G$ we also have 
\be \label{23}
O^{(G)}(f)\approx O^{(D)} \left( \left[ \exp(\beta_I \{(\tilde{M}^{-1})^{IJ} S_J,.\})\cdot
f]_{\beta=-(G-D)} \right] \right) \text{,}
\ee
with $\tilde{M}_{IJ}=\{S_,G_J\}$. Notice that the argument of $O^{(D)}$ on 
the right hand side of (\ref{23}) is not gauge invariant and that it is 
the gauge transform of $f$ with respect to the weakly Abelian constraints 
$\tilde{S}_I=[\tilde{M}^{-1}]^{IJ} S_J$ from the gauge $G=0$ to the gauge 
$D=0$ as desired.\\
5.\\
An important final comment concerns the dynamics of the theory (we consider
for simplicity only one pair of second class constraints but the same
discussion applies, with more notational load, to the general case):\\
Suppose first that $H'=H_0+\lambda^I S_I$, that is $H_0$ is not constrained
to vanish. From the point of view of the second class system the Hamiltonian
that drives the dynamics of the system is $H$ or equivalently $\tilde{H}$
via the Dirac bracket evaluated on the constraint surface of the second
class system $\overline{\overline{{\cal M}}}$, that
is
\be \label{24}
\dot{f}_{|S=D=0}=[\{\tilde{H},f\}^\ast_{S,D}]_{S=D=0}
=\{H_{|S=D=0},f_{|S=D=0}\}^\ast_{S,D} \text{.}
\ee
On the other hand, from the point of view of the first class system, the
Hamiltonian is $\tilde{H}$ which acts on gauge ($S$-) invariant functions
which we write in the form $F=O^{(D)}(f)$ on the constraint surface of the
first class system $\overline{{\cal M}}$, that is
\be \label{25}
\dot{F}_{|S=0}=\{\tilde{H},F\}_{|S=0}=\{O^{(D)}(H),O^{(D)}(f)\}_{|S=0}
=O^{(D)}(\{H,f\}^\ast_{S,D})_{|S=0} \text{.}
\ee
Comparing (\ref{24}) and (\ref{25}) we see that the time evolutions are
isomorphic when mapping $f_{|{S=D=0}}$ to $O^{(D)}(f)_{|S=0}$.\\
Now we consider the case that $H_0=C$ itself is constrained to vanish.
Then also the Hamiltonian $\tilde{H}$ is constrained to vanish from the
point of view of the second class system since is a linear combination of the
three constraints $C,S,D$. Now the following subtlety arises: From the
point of view of the first class system, the Hamiltonian $\tilde{H}$
is {\it not} constrained to vanish because the first class system is only
subject to the constraints $C, S$. But this would clearly be wrong: The
first class system would only have the constraint $S$ and this would lead
to a different dimensionality of the reduced phase space than in the second
class system. The correct point of view is the following: The second class
system is equivalently described by the three types of constraints
$\tilde{H},S,D$ of which $\tilde{H}$ constitutes a first class set of
constraints while $(S,D)$ constitute a second class system of constraints.
From the point of view of the first class system we just forget about the
$D$ constraints and instead consider the first class constraint system
$\tilde{H},S$. The counting of physical number of degrees of freedom is
now correct again because both first class constraints $\tilde{H},S$ count
twice in the first class system while in the second class system
$\tilde{H},S,D$ only $\tilde{H}$ counts twice and $S,D$ only count once.
This also makes sure that there is no true Hamiltonian in both schemes.
To compare the observables from both points of view, let
$S_1:=S,\;S_2:=\tilde{H},\;D_1:=D,\;D_2:=G$ where the gauge fixing condition
$G$ is chosen in such
a way that the matrix with entries $M_{IJ}=\{S_I,D_J\}$ is non singular.
It is easy to see that the second class system $(S_I,D_I)$ is of the
type to which gauge unfixing applies and the discussion proceeds from here
just as in the general case.

\section{Application of Gauge Unfixing to Gravity}
\label{section_gugr}

We now want to apply the ideas of gauge unfixing to higher dimensional General Relativity and start with the Hamiltonian system derived in section \ref{section_ham2}. The second class constraints are given by $S^{ab}_{\overline{M}} \approx D^{ab}_{\overline{M}} \approx 0$. As we pointed out in section \ref{sec:ConstraintAnalysis}, the constraints are not independent and the Dirac matrix is not invertible. We will neglect this fact for the moment and remark that we can deal with it using the independent sets of constraints discussed above. We remark that gauge unfixing has been applied previously to $2+1$-dimensional linearised massive gravity \cite{AriasGaugeInvarianceAnd}. 

The general discussion of the previous section suggests that the simplicity invariant extension
of the Hamiltonian constraint involves an infinite series which is beyond any analytical
control already at the classical level.
Luckily, the Dirac matrix depends only on $\pi^{aIJ}$ and therefore commutes with the BF-simplicity constraint. Hence repeated commutators acting
on functions that depend polynomially on $A$ vanish beyond the order of the polynomial.
We calculate explicitly
\be  \tilde{H} = H - \frac{1}{2} D^{ab}_{\overline{M}} \; \left( F^{-1} \right) \m_{ab}^{\overline{M}} \, \m_{cd}^{\overline{N}} \; D^{cd}_{\overline{N}} \text{,} \ee 
where terms up to the second order contributed, since $H$ is quadratic in the connection. The effect of the extra term in the Hamiltonian can be seen when solving the simplicity constraint and reducing the theory to the ADM variables. When doing the calculation (\ref{reduction_h}), we have to use $D \sim (F \bar{K}^T)^{aIJ}=0$ to eliminate a term proportional to $\bar{K}^T_{aIJ} F^{aIJ, bKL} \bar{K}^T_{bKL}$. This is not necessary any more because the additional $-1/2 D F^{-1} D$ precisely counters this term. 

The Gau{\ss} and diffeomorphism constraints only obtain extra terms proportional to the BF-simplicity constraints which can be neglected in the first class theory. We can use the projector identities to calculate the new constraint algebra
\begin{eqnarray}
 \left\{ \tilde{G}, \tilde{G} \right\} & = & \tilde{G} + S \text{,}\\
 \left\{ \tilde{G}, \tilde{\mathcal{H}}^a \right\} & = & S \text{,}\\
 \left\{ \tilde{G}, \tilde{\mathcal{H}} \right\} & = &  S \text{,}\\
 \left\{ \tilde{\mathcal{H}}_a, \tilde{\mathcal{H}}_b \right\} & = & \tilde{\mathcal{H}}_a + \tilde{G} + S \text{,}\\  
 \left\{ \tilde{\mathcal{H}}_a, \tilde{\mathcal{H}} \right\} & = & \tilde{\mathcal{H}} + \tilde{G} + S\text{,} \\
 \left\{ \tilde{G}, S \right\} & = & S \text{,}\\ 
 \left\{ \tilde{\mathcal{H}}_a, S \right\} & = & S \text{,}\\ 
 \left\{ \tilde{\mathcal{H}}, S \right\} & = & 0 \text{.}
\end{eqnarray} 
By construction it closes without the $D$ constraint and displays a first class structure. 

Concerning gauge invariant phase space functions, it is clear that a vanishing commutator with the BF-simplicity constraint does not constrain the dependence on $\pi^{aIJ}$. Additionally, these functions may only depend on the simplicity invariant extension of $A_{aIJ}$ which is 
given explicitly by
\be \tilde{A}_{aIJ} =  A_{aIJ} + D^{cd}_{\overline{N}} \; \left( F^{-1} \right) \m_{cd,}^{\overline{N}} \m_{ab}^{\overline{M}} \; \epsilon_{\overline{M} IJKL } \pi^{bKL} \text{,} \ee
since $A_{aIJ}$ changes under simplicity gauge transformations as
\be \delta^S A_{aIJ} := \left\{ A_{aIJ}, S^{bc}_{\overline{M}}[c_{bc}^{\overline{M}}] \right\} = c_{ab}^{\overline{M}} \epsilon_{IJKL\overline{M}} \pi^{bKL} \text{.} \ee

We still have to give a sense to $\left( F^{-1} \right) \m_{cd,}^{\overline{N}} \m_{ab}^{\overline{M}}$. As we have shown in section \ref{section_ham2}, it is enough to consider only a subspace of Lagrange multipliers for the BF-simplicity and $D^{ab}_{\overline{M}}$ constraints parametrised by the projected test functions
\be  d_{ab}^{\overline{M}} = \bar{d}_{(a|IJ} \pi_{b)KL} \epsilon^{IJKL \overline{M}} \text{.}  \ee 
On this subspace, $F^{aIJ,bKL}$ was shown to be invertible. We therefore make the Ansatz
\be  \left( F^{-1} \right) \m_{cd,}^{\overline{N}} \m_{ab}^{\overline{M}} = \gamma \epsilon^{EFGH \overline{N}} \pi_{(c|EF} \left(F^{-1}\right)_{d)GH,(a|AB} \pi_{b)CD} \epsilon^{ABCD\overline{M}}  \ee 
for some constant $\gamma$, where 
\be  \left( F^{-1} \right)_{aIJ,bKL} := \frac{s}{4 (D-1)} \pi_{aAC} \pi_{bBD} \left(\pi^{cEC} \pi_{cE} \m^D -s \eta^{CD} \right) \left( \eta^{AB} \eta^{K[I} \eta^{J]L} - 2 \eta^{LA} \eta^{B[I} \eta^{J]K} \right)  \ee 
only depends on $\pi^{aIJ}$ and reduces to the correct expression on the simplicity constraint surface when contracted in the above equation. Insertion into $\tilde{A}$ yields
\be \gamma = \frac{1}{4 (D-1)^2((D-3)!)^2}  \ee 
when demanding that $\tilde{A}$ is independent of $D$, i.e. that the $\bar{K}^T_{aIJ}$ term 
is cancelled. 
Since all simplicity invariant phase space functions are arbitrary functions  of $\tilde{A}_{aIJ}$ and $\pi^{aIJ}$, we have shown that the proposed expression for $\left( F^{-1} \right) \m_{cd,}^{\overline{N}} \m_{ab}^{\overline{M}}$ yields the desired results. 
This can of course also be obtained by direct inversion of the projected version of the matrix $F$. This way we obtain a connection formulation for gravity in $D+1>3$ without second class constraints. Notice however that the observables (with respect to the simplicity constraint) 
$(\tilde{A},\pi)$ have complicated Poisson brackets, only the brackets of the canonical 
pair $(A,\pi)$ are simple, therefore suggesting a Dirac quantisation approach (quantisation
at the kinematical level).\\
\\
Let us summarise and compare with the connection formulation in $D+1=4$:\\ 
1.\\
On the surface where the simplicity constraint vanishes, $\pi^{aIJ} = 2n^{[I}E^{a|J]}$, we can 
describe the situation more explicitly. From the above formula it is obvious that both $ n^I \delta^S A_{aIJ} = 0$ and $E^{aI} \delta^S A_{aIJ} = 0$, since we always may choose $c_{[ab]}^{\overline{M}} = 0$. Thus, when decomposing the connection $A_{aIJ} = \Gamma_{aIJ} + \bar{K}_{aIJ} + 2 n_{[I}K_{a|J]}$ into hybrid spin connection and rotational (i.e. transversal) and boost (i.e longitudinal) components of hybrid contorsion $K_{aIJ}$, we find that the simplicity constraint generates on-shell gauge transformations of the trace free part of the rotational (transversal) components of the SO$(1,D)$ hybrid contorsion $\bar{K}_{aIJ}^T$. As we have seen in equation (\ref{eq:TraceRotKomp}), the remaining trace component of the rotational part $E^{aI} \bar{K}_{aIJ}$ is proportional to the boost part of the 
Gau{\ss} constraint and vanishes if $n_I G^{IJ} = 0$ holds. In total, we find that observables in this connection theory may not depend on the value of the rotational components of the SO$(1,D)$ hybrid contorsion at all. The whole physical information contained in the connection is encoded in the boost components of the contorsion, which becomes conjugate to the vielbein after solving the simplicity constraint. Therefore, when removing the boost gauge freedom by choosing 
the time gauge, there is no physical information left in the SO$(D)$ connection.\\ 
2.\\
In $D+1=4$, this formulation therefore differs from the formulation in terms of real Ashtekar variables considered in \cite{SaHamiltonianAnalysisOf}, which remains a connection formulation also after imposing the time gauge. This is achieved by mixing boost and rotational components of the connection using the total antisymmetric tensor, i.e. $\m^{(\beta)}A_{ajk} = A_{ajk} - \beta \epsilon_{0ijk} A_{a0i}$, to ``rotate'' physical degrees of freedom into the rotational components of the connection. Thus, this procedure exploits a peculiarity of dimension $D+1=4$, and therefore is not possible in any other dimension. In $D+1=4$ it is possible to arrive at this connection formulation in terms of 
$^{(\beta)}A$ also by enlarging the ADM phase space instead of starting from the Holst action 
\cite{SaHamiltonianAnalysisOf}. This turns out to be true also for the new connection formulation derived above
as we have shown in our companion paper \cite{BTTI}.
Following this route allows for the introduction of a free parameter similar to the Barbero-Immirzi parameter, but the transformation made to obtain the connection is very different in nature since there is no mixing of boost and rotational parts.\\ 
3.\\
The internal gauge group in the case of Lorentzian external signature is SO$(1,D)$, for which the techniques developed in \cite{AshtekarRepresentationsOfThe, AshtekarRepresentationTheoryOf} are not available. It is therefore desirable to obtain a formulation with a compact internal gauge group as is the case in $D+1=4$. The idea, similar
to the situation in 
$D+1=4$, is to add a correction term to the Euclidean Hamiltonian constraint which changes the external signature after reduction to ADM variables. This has been achieved in \cite{BTTI}
where we derived the formulation presented in this section and solved the task to obtain a compact gauge group formulation even for Lorentzian signature. In contrast to $D+1=4$ this 
is not achieved by choosing a gauge in a formulation coming from a Lorentzian gauge theory action but
by simply choosing a compact gauge group extension of the Lorentzian ADM formulation.
This formulation therefore cannot be obtained from a Lorentzian connection 
formulation.

\section{Conclusions}

\label{section_cr}

We have shown that General Relativity in dimensions $D+1 \geq 3$ can be written as a gauge theory with either SO$(D+1)$ or SO$(1,D)$ as the internal gauge group. We derived this 
result both from a canonical analysis of the Palatini Lagrangian in this paper and 
from an extension of the ADM phase space using Hamiltonian methods in \cite{BTTI}. In both 
cases, the simplicity constraints in addition to the Gau{\ss} constraints played a key role
in order to match the correct number of degrees of freedom. However, there are two differences:\\
First, the Lagrangian methods are restricted to a matching of spacetime and internal 
signature, that is, we get necessarily SO$(1,D)$ for Lorentzian General Relativity. There 
is no room for an Immirzi like parameter (without a topological term in the Lagrangian 
which we did not consider here since in higher dimensions such a term is not known 
in explicit form depending on $D$). 
For the Hamiltonian methods there is no such restriction, in particular we can have SO$(D+1)$
for Lorentzian General Relativity and arbitrary Immirzi like Parameter. The price to pay is that this connection formulation does not 
have a Lagrangian origin, on the other hand, Loop Quantum Gravity methods are available
for quantisation because the gauge group is compact which is a huge advantage.\\
Secondly, as is well known, the Palatini action gives rise to second class constraints and 
an associated non trivial Dirac bracket with respect to which the connection is not 
commuting (see however \cite{AlexandrovSU(2)LoopQuantum, AlexandrovCriticalOverviewOf}) thus forbidding a connection formulation (the connection cannot act as a 
multiplication operator in the quantum theory) \cite{AlexandrovSO4CCovariantAshtekar}. On the other hand, the Hamiltonian extension
of the ADM phase space does not know about any Dirac bracket and starts with a connection
that is Poisson commuting and with only first class constraints. 

This second observation raises the questions how these two descriptions can possibly 
be simultaneously valid? The answer to this puzzle is that (for matching spacetime and internal
signature) the Hamiltonian description can be obtained from the Lagrangian description by  gauge unfixing the second class theory into the first class theory. The secondary second class 
constraint partner to the primary second class simplicity constraint thereby gets removed and 
gets built into the Hamiltonian constraint in order to achieve the first class property. 
From the Hamiltonian perspective this secondary second class constraint precisely arises as a 
counterterm that removes a piece in the ``natural'', primary Hamiltonian constraint (expressed in terms of curvature) absent in the correct first class Hamiltonian constraint.     

The description at which we arrived in this paper is therefore very different from the platform 
from which Lorentzian Loop Quantum Gravity in $D+1=4$ starts. There is no time gauge 
and therefore the gauge group is SO$(4)$ rather than SU$(2)$. The price to pay are the additional
simplicity constraints and a more complicated Hamiltonian constraint. For $D+1=4$ 
this description is therefore more complicated than the standard description and thus 
less useful. However, for $D+1>4$ we do not know of any other gauge theory formulation 
of General Relativity with 1. compact  gauge group and 2. only first class constraints and 3.
standard symplectic structure with Poisson commuting connections. We will address 
the differences in the quantum theories between the present formulation and standard LQG
in a future publication. We expect that there will be close relations with the way the 
simplicity constraint is treated in the EPRL Spin Foam model \cite{EngleTheLoopQuantum, LivineNewSpinfoamVertex, EngleFlippedSpinfoamVertex, EngleLoopQuantumGravity, FreidelANewSpin} as well as with 
projected spin networks \cite{LivineProjectedSpinNetworks}. In fact, it is very likely that the present formulation 
will present the bridge between the kinematical Hilbert space of standard LQG and 
the boundary Hilbert space of current Spin Foam models. The reason is that the simplicity constraints 
will require the SO$(4)$ representations and intertwiners to be ``simple'', i.e. there will be some 
relation between the left and right handed spin labeling an irreducible SO$(4)$ representation
precisely as in current Spin Foam models so that effectively one deals with an SU$(2)$ theory 
as in standard LQG. What was missing so far is the interpretation of the connection in terms of which that effective SU$(2)$ theory is formulated as compared to the Ashtekar-Barbero connection. Since Spin Foam models start from the 
Palatini (or equivalently the Plebanski) action as in our case, it is very suggestive to assume 
that the Spin Foam boundary Hilbert space is formulated in terms of our formulation with the twist 
of the gauge unfixing procedure as otherwise one could not have a connection 
representation.\\
\\
In our companion papers, we will extend the methods developed for $3+1$ Loop Quantum Gravity to our $D+1$ theory and show that it can be quantised analogously \cite{BTTIII}, extend the treatment to standard matter \cite{BTTIV}, discuss possible solution strategies for the simplicity constraint \cite{BTTV}, and finally develop tools allowing the loop quantisation of a large class of Supergravity theories \cite{BTTVI, BTTVII}. 
\\
\\
\\
{\bf\large Acknowledgements}\\
NB and AT thank Yuriy Davygora, Muxin Han,  and Johannes Tambornino for numerous discussions as well as the AEI for hospitality during parts of this work. NB and AT thank the Max Weber-Programm,  the German National Merit Foundation, and the Leonardo-Kolleg of the FAU Erlangen-N\"{u}rnbeg for financial support. NB further acknowledges financial support by the Friedrich Naumann Foundation. The part of the research performed at the Perimeter Institute for Theoretical Physics was supported in part by funds from the Government of Canada through NSERC and from the Province of Ontario through MEDT.

\newpage

\begin{appendix}

\setcounter {equation} {0}

 \renewcommand\theequation{\thesection .\arabic{equation}}

\section{The Lie algebra so$(1,D)$}
\label{appendix_liealgebra}
In this appendix, we generalise a so$(1,3)$ structure constant identity given in \cite{PeldanActionsForGravity}. In our notation, 
\be  \left( T_{AB} \right)^I \m_J = \eta^I \m_{[A} \eta_{B]J} \ee 
denotes the generators of so$(1,D)$ in the fundamental representation. The \textit{antisymmetric} index pair $AB$ labels the $D(D+1)/2$ generators, $I$ and $J$ are matrix indices, also antisymmetric. In the following, a generator $T_{AB}$ will always have a label, but the matrix indices will be mostly suppressed. Insertion of the definitions shows that the generators satisfy the usual Lorentz algebra
\be  \left[ T_{AB}, T_{CD} \right]^I \m_J = 2 \eta_{A][C} \left(T_{D][B} \right)^I \m_J =: f_{AB,CD,} \m^{EF} \left( T_{EF} \right)^I \m_J  \ee 
with
\be f_{AB,CD,EF} = -2 \eta_{B][C}\eta_{D][E}\eta_{F][A} = -2\text{Tr} \left(T_{AB} T_{CD} T_{EF} \right) \text{.} \label{eqn_etaf} \ee
We further define the Cartan-Killing metric
\be  q_{IJ,KL} = \eta_{I[K} \eta_{L]J} ~ \Leftrightarrow ~ -\text{Tr} \left( T_{AB} T_{CD} \right) = \left( T_{AB} \right)^{IJ} q_{IJ,KL}\left( T_{CD} \right)^{KL}\ee 
and the object
\be  (q^{*\overline{M}})_{IJ,KL} = \frac{1}{2} \epsilon_{IJKL}^{~~~~~~\overline{M}} \ee 
defining the dual
\be T^{*\overline{M}}_{AB} = (q^{*\overline{M}})_{AB,} \m^{CD} T_{CD} \ee 
generators. We note that self-duality is a concept reserved for $3+1$ dimensions. 

These definitions lead us to the main result of this appendix:
\be \label{eqn_ff}  f_{AB,CD,IJ} f_{EF,GH,} \m^{IJ} = \frac{1}{2} q_{AB,EF} q_{GH,CD}-\frac{\eta_{\overline{M}\overline{N}}}{2(D-3)!}(q^{*\overline{M}})_{AB,EF}(q^{*\overline{N}})_{GH,CD} - EF \leftrightarrow GH\text{.} ~~\ee 
It can be proven by carefully inserting the definitions and writing out explicitly each term.

\section{Introduction of the Barbero-Immirzi Parameter in $3+1$ Dimensions}
\label{sec:Immirzi}

In the special case of $3+1$ dimensions, it is straight forward to introduce the Barbero-Immirzi parameter $\gamma$: Perform the canonical analysis of the Holst action without time gauge (given in, e.g., \cite{SaHamiltonianAnalysisOf}) and use the method of gauge unfixing in complete analogy to the treatment in this paper. The matrix $F$, which has to be inverted for gauge unfixing, is very simple in this case, given by $F^{ab, cd} = 2 q^2 q^{a[b} q^{c]d}$ \cite{SaHamiltonianAnalysisOf}. Following these lines, we obtain a connection formulation with (first class) quadratic simplicity constraints and gauge group SO$(3,1)$, which reduces to the Ashtekar-Barbero formulation after solving the Simplicity and boost Gau{\ss} constraints. (Another straight-forward calculations shows that the same procedure gives a possible solution to the open issue $(i)$ in \cite{WielandComplexAshtekarVariables}). For quantisation purposes, it again would be nice to be able to work with the compact gauge group SO$(4)$ instead of the Lorentz group. Moreover, the linear simplicity constraint, which is introduced in any dimension $D+1$ in \cite{BTTVI}, is favoured in $3+1$ dimensions, since the quadratic simplicity constraint allows for unphysical solutions, usually called the topological sector. Last but not least, a formulation with Barbero-Immirzi parameter and linear simplicity constraints maximally mimics Spin Foams. In this appendix, we will show by extending ADM phase space that both, the formulations with flipped signature and with linear simplicity constraints, exist.

\subsection{Flipped Signature and Quadratic Simplicity Constraints}
\label{app:Quadratic}

We start with the formulation given in \cite{BTTVI} with variables $\left\{K_{aIJ}, \pi^{bKL}\right\}$. In $3+1$ dimensions, we can introduce for real Barbero-Immirzi parameter $\gamma$, $\gamma^2 \neq \zeta$,
\be
(\overset{(\gamma)}{\mathcal{M}})^{IJ}\m_{KL} :=  \eta^{[I}_K \eta^{J]}_L + \frac{1}{2\gamma} \epsilon^{IJ}\m_{KL} ~~\text{and} ~~ (\overset{(\gamma)}{\mathcal{M}}\m^{-1})^{IJ}\m_{KL} :=  \frac{\gamma^2}{\gamma^2-\zeta} \left(\eta^{[I}_K \eta^{J]}_L - \frac{1}{2\gamma} \epsilon^{IJ}\m_{KL}\right)\text{.}
\ee
It is obvious that the transformation to the canonical pair of variables defined by 
\ba
\underset{(\gamma,\beta)}{K}\m_{aIJ} &:=& \beta (\overset{(\gamma)}{\mathcal{M}}\m^{-1})\m_{IJ}\m^{KL} K_{aKL} := \beta \overset{(\gamma)}{\mathcal{M}}\m^{-1} K_{aIJ} \text{,} \\
\overset{(\gamma,\beta)}{\pi} \m^{aIJ} &:=& \frac{1}{\beta} (\overset{(\gamma)}{\mathcal{M}}) \m^{IJ}\m_{KL} \pi^{aKL} := \frac{1}{\beta} \overset{(\gamma)}{\mathcal{M}}\pi^{aIJ} \text{,}
\ea
is canonical. Note that we introduced a second free parameter $\beta$ coming from a constant rescaling, which already appeared in \cite{BTTI}. To obtain a connection formulation, we the would like to use the canonical pair of variables given by $\left\{A_{aIJ} := (\Gamma + \underset{(\gamma,\beta)}{K})\m_{aIJ}, \overset{(\gamma,\beta)}{\pi} \m^{aIJ}\right\}$. We will prove in the following that these variables are indeed a valid extension of the ADM phase space. 

For later convenience, we introduce the notations
\ba
\overset{(\gamma, \beta)}{\pi}\m_{aIJ} &:=& \frac{1}{q} q_{ab} \overset{(\gamma, \beta)}{\pi}\m^{b}\m_{IJ} \text{,} \nonumber \\
\pi^{aIJ} &:=& \beta \cdot (\overset{(\gamma)}{\mathcal{M}}\m^{-1})^{IJ}\m_{KL} \overset{(\gamma,\beta)}{\pi}\m^{aKL}\text{,} \nonumber \\
\underset{(\gamma,\beta)}{\pi}\m^{aIJ} &:=& \beta^2 \cdot (\overset{(\gamma)}{\mathcal{M}}\m^{-1})^{IJ}\m_{KL} (\overset{(\gamma)}{\mathcal{M}}\m^{-1})^{KL}\m_{MN}\overset{(\gamma,\beta)}{\pi}\m^{aMN} \nonumber \text{,}
\ea
where in the first line $\frac{1}{q} q_{ab}$ has to be understood as a function of $\overset{(\gamma,\beta)}{\pi}\m^{aIJ}$ as given in (\ref{eq:Metric}). Moreover, note that in $3+1$ dimensions the expression for the hybrid spin connection (cf. \cite{BTTI}) can be simplified to
\be
\Gamma_{aIJ} := \zeta \pi_{b[I| K} \nabla_a \pi^{b}\m_{J]}\m^{K} := \zeta \left( \pi_{b[I| K} \partial_a \pi^{b}\m_{J]}\m^{K} + \pi_{b[I| K} \Gamma_{ac}^b \pi^{c}\m_{J]}\m^{K} \right)\text{,} \label{eq:SpinConnection}
\ee
where $\Gamma^b_{ac}$ denotes the Christoffel symbols and therefore, $\nabla_a$ is the covariant derivative annihilating $q_{ab}$. The ADM variables, expressed in terms of $A_{aIJ}$ and $\overset{(\gamma,\beta)}{\pi} \m^{aIJ}$, are given by
\ba
2 \zeta q q^{ab} &:=& \pi^{aIJ} \pi^{b}_{IJ} = \overset{(\gamma,\beta)}{\pi}\m^{aIJ} \underset{(\gamma,\beta)}{\pi}\m^{b}_{IJ}  \label{eq:Metric} \text{,}\\
K^{ab} &:=& - \frac{s}{4 \sqrt{q}} \overset{(\gamma,\beta)}{\pi}\m^{(b| IJ} q^{a)c} \left(A-\Gamma \right)_{cIJ} \label{eq:ExtrinsicCurvature} \text{,}\\
P^{ab} &=& -s\sqrt{q}\left(K^{ab} - q^{ab}K_c\m^c \right)=\frac{1}{4} \left(q^{c(a}\overset{(\gamma,\beta)}{\pi}\m^{b)IJ} - q^{ab} \overset{(\gamma,\beta)}{\pi}\m^{cIJ}\right) \left(A-\Gamma \right)_{cIJ} \label{eq:MetricConjugateMomentum} \text{.}
\ea
Using the above equations (\ref{eq:Metric}, \ref{eq:MetricConjugateMomentum}), we find for the ADM constraints
\ba
 \mathcal{H}_a &=& -2 q_{ac} \nabla_b P^{bc} \nonumber \\
                            &=& -\frac{1}{2} \nabla_b \left( \left( A-\Gamma \right)_{aIJ} \overset{(\gamma,\beta)}{\pi}\m^{bIJ} - \delta_a^b \left( A-\Gamma \right)_{cIJ}\overset{(\gamma,\beta)}{\pi}\m^{cIJ} \right) \label{eq:Diffeo} \text{,} \\
 \mathcal{H} &=& - \left[ \frac{s}{\sqrt{q}} \left( q_{ac} q_{bd} - \frac{1}{2} q_{ab} q_{cd} \right) P^{ab} P^{cd} + \sqrt{q} R \right]  \nonumber \\
                       &=& - \frac{s}{8\sqrt{q}} \left( \overset{(\gamma,\beta)}{\pi}\m^{[a|IJ}\overset{(\gamma,\beta)}{\pi}\m^{b]KL} \left(A-\Gamma \right)_{bIJ} \left(A-\Gamma \right)_{aKL}\right) - \sqrt{q} R(\overset{(\gamma,\beta)}{\pi}) \label{eq:Hamilton} \text{.}
\ea
In order to have the right number of physical degrees of freedom, the Gau{\ss} and quadratic simplicity constraints
\ba
G^{IJ} &:=& D_a\overset{(\gamma,\beta)}{\pi}\m^{aIJ} \approx 2 \left(A-\Gamma \right)_{a}\m^{[I}\m_K \overset{(\gamma,\beta)}{\pi}\m^{aK|J]} \label{eq:Gauss} \text{,} \\
S^{ab} &:=& \frac{1}{4} \epsilon^{IJKL}  \pi^a_{IJ}  \pi^b_{KL} \text{,} \label{eq:Simplicity}
\ea
are introduced. In $3+1$ dimensions, the quadratic simplicity constraint has additional solutions which lead to a theory not corresponding to General Relativity. We will exclude this sector by hand. In section \ref{sec:Lin}, we will introduce the linear version of the constraint like in \cite{BTTVI}, which does not have this additional solution sector. 
Using the (non-vanishing) Poisson brackets
\be
\left\{A_{aIJ}(x), \overset{(\gamma,\beta)}{\pi}\m^{bKL} (y)\right\} := 2 \delta_a^b \delta_{[I}^{K} \delta_{J]}^L \delta^3(x-y) \label{eq:PoissonBrackets} \text{,}
\ee
one can check that $\Gamma_{aIJ}$ given in (\ref{eq:SpinConnection}) transforms as a connection under the action of the Gau{\ss} constraint, and therefore (\ref{eq:ExtrinsicCurvature}, \ref{eq:MetricConjugateMomentum}, \ref{eq:Diffeo}, \ref{eq:Hamilton}) are invariant under gauge transformations. Since the matrix $\overset{(\gamma)}{\mathcal{M}}$ is built from intertwiners, (\ref{eq:Metric}) and (\ref{eq:Simplicity}) are gauge invariant by inspection. Simplicity invariance of (\ref{eq:MetricConjugateMomentum}, \ref{eq:Diffeo}, \ref{eq:Hamilton}) follows from
\ba
\left\{K^{ab}(x), S^{cd}(y) \right\} &=& - \frac{s}{4 \sqrt{q}} \overset{(\gamma,\beta)}{\pi}\m^{(b| IJ} q^{a)e}(x)  \left\{A_{eIJ}(x),  \frac{1}{4} \epsilon^{KLMN}  \pi^c_{KL}(y) \pi^d_{MN}(y) \right\} \nonumber \\
&=& - \frac{s}{4 \sqrt{q}}  \epsilon^{IJKL}  \beta \overset{(\gamma)}{\mathcal{M}}\m^{-1}\overset{(\gamma,\beta)}{\pi}\m^{(b| IJ} q^{a)(c} \pi^{d)}_{KL} ~ \delta^3(x-y) \nonumber \\
&=& -\frac{s}{2\sqrt{q}} \left( S^{b(d} q^{c)a} + S^{a(d} q^{c)b}\right) ~ \delta^3(x-y) \approx 0 \text{.} 
\ea
What remains to be checked is if the ADM Poisson brackets are reproduced on the new phase space, which will by construction imply that the constraint algebra closes. The following Poisson brackets will be helpful in the sequel:
\ba
\left\{A_{aIJ}(x), q^{bc}(y)\right\} &=& \zeta \delta^3(x-y) \left( \frac{2}{q} \delta_{a}^{(b} \underset{(\gamma,\beta)}{\pi}\m^{c)}\m_{IJ} - q^{bc}  \underset{(\gamma,\beta)}{\pi}\m_{aIJ} \right) \label{eq:qOben} \text{,}\\
\left\{A_{aIJ}(x), q_{bc}(y)\right\} &=& -\zeta \delta^3(x-y) \left( 2 q_{a(b} \underset{(\gamma,\beta)}{\pi}\m_{c)IJ} - q_{bc}  \underset{(\gamma,\beta)}{\pi}\m_{aIJ} \right) \label{eq:qUnten} \text{,} \\
\left\{A_{aIJ}(x), q (y)\right\} &=& \zeta \delta^3(x-y) q \cdot \underset{(\gamma,\beta)}{\pi}\m_{aIJ} \label{eq:Detq} \text{,} \\
 \overset{(\gamma,\beta)}{\pi}\m^{c IJ}(x) \left\{A_{aIJ}(x), \pi_{b}\m^{KL}(y)\right\} &\approx& \delta^3(x-y) \left( \frac{2}{q} q_{ab} \bar{\eta}_{[M}^K \bar{\eta}_{N]}^L - \zeta \pi_a\m^{KL} \pi_{bMN} \right) \beta (\overset{(\gamma)}{\mathcal{M}}\m^{-1})\m^{MN}\m_{IJ} \overset{(\gamma,\beta)}{\pi}\m^{c IJ} \nonumber \\
 &\approx& - 2 \delta^3(x-y) \pi_a\m^{KL} \delta_b^c \label{eq:piUnten} \text{.}
\ea
The brackets
\be
\left\{q_{ab}, q_{cd}\right\}_{(A,\overset{(\gamma, \beta)}{\pi})} = 0~~~\text{and}~~~ \left\{q_{ab}, P^{cd}\right\}_{(A,\overset{(\gamma, \beta)}{\pi})} = \delta_{(a}^c\delta_{b)}^d
\ee
are easily verified. The remaining Poisson bracket
\begin{alignat}{3}
 &\left\{P^{ab}[A_{ab}], P^{cd}[B_{cd}]\right\}_{(A,\overset{(\gamma, \beta)}{\pi})}  \nonumber \\
=& \int_{\sigma} d^3x \int_{\sigma} d^3y \left[ \left(\frac{1}{2} A_{ab} q^{a[e}  \overset{(\gamma,\beta)}{\pi}\m^{b]IJ}\right)(x) \left\{A_{eIJ}(x), \left(q^{c[f}  \overset{(\gamma,\beta)}{\pi}\m^{d]KL}\right)(y) \right\} \left(\frac{1}{2} B_{cd} (A-\Gamma)_{fKL}\right)(y)\right] \nonumber \\
 &  - [A \leftrightarrow B]  \label{eq:SecondLine} \\
 & + \int_{\sigma} d^3x \int_{\sigma} d^3y \left[ \left(\frac{1}{2} A_{ab} q^{a[e}  \overset{(\gamma,\beta)}{\pi}\m^{b]IJ}\right)(x) \left\{A_{eIJ}(x),  (-\Gamma_{fKL})(y)\right\} \left(\frac{1}{2} B_{cd} q^{c[f}  \overset{(\gamma,\beta)}{\pi}\m^{d]KL} \right)(y) \right] \nonumber \\
 & - [A \leftrightarrow B]  \label{eq:ThirdLine} 
\end{alignat}
is much harder and therefore will be discussed in more detail. Here, $A_{ab}$ and $B_{cd}$ are test fields of compact support, which we can choose symmetric w.l.o.g., since $P^{ab}$ is symmetric by definition. The second line (\ref{eq:SecondLine}) and third line (\ref{eq:ThirdLine}) of the above equation vanish independently. For (\ref{eq:SecondLine}), we find using (\ref{eq:qOben})
\be
(\ref{eq:SecondLine}) = \hdots = \frac{1}{4} \int_{\sigma} d^3x ~ A_{ab} B_{cd} q^{ac}  \overset{(\gamma,\beta)}{\pi}\m^{[d|IJ} q^{b]e} (A-\Gamma)_{eIJ} \propto \bar{G}^{IJ}[\hdots] \text{,} \nonumber
\ee
which vanishes if the (rotational part of the) Gau{\ss} constraint holds. Before we proceed, we define $\alpha_f\m^e := \frac{1}{4} A_{ab} \left( q^{e(a} \delta^{b)}_f - q^{ab} \delta_f^e\right)$ and  $\beta_h\m^g := \frac{1}{4} B_{cd} \left( q^{g(c} \delta^{d)}_h - q^{cd} \delta_h^g\right)$ and check that $\alpha_{[ef]} = 0=\beta_{[gh]}$. Then, we find for the third line (skipping ``$- [A \leftrightarrow B]$" for a moment)
\begin{alignat}{3}
(\ref{eq:ThirdLine}) =&  \int_{\sigma} d^3x \int_{\sigma} d^3y ~ \alpha_f\m^e \overset{(\gamma,\beta)}{\pi}\m^{f IJ}(x)\left\{A_{eIJ}(x),  (- \zeta) \pi_{bKM} \left(\nabla_g \pi^{b}\m_{L}\m^{M}\right) (y)\right\} \beta_h\m^g \overset{(\gamma,\beta)}{\pi}\m^{hKL}(y) \nonumber \\
=&  - \zeta \int_{\sigma} d^3x \int_{\sigma} d^3y ~ \alpha_f\m^e \overset{(\gamma,\beta)}{\pi}\m^{f IJ}(x) \Big[ \left\{A_{eIJ}(x), \pi_{bKM}(y) \right\}  \left(\nabla_g \pi^{b}\m_{L}\m^{M}\right)\beta_h\m^g \overset{(\gamma,\beta)}{\pi}\m^{hKL}(y) \nonumber \\
  & \hspace{40mm} - \left\{A_{eIJ}(x), \pi^{b}\m_{L}\m^{M} (y)\right\}  \nabla_g\left(\pi_{bKM} \beta_h\m^g \overset{(\gamma,\beta)}{\pi}\m^{hKL}\right)(y) \Big] \label{eq:IIA} \\
 & - \zeta \int_{\sigma} d^3x \int_{\sigma} d^3y ~ \alpha_f\m^e \overset{(\gamma,\beta)}{\pi}\m^{f IJ}(x) \left\{A_{eIJ}(x),  \Gamma_{ga}^b(y)\right\}  \pi_{bKM} \pi^{a}\m_{L}\m^{M} \beta_h\m^g \overset{(\gamma,\beta)}{\pi}\m^{hKL}(y) \text{.}\hspace{15mm}  \label{eq:IIB}
\end{alignat}
Again, (\ref{eq:IIA}) and (\ref{eq:IIB}) vanish separately. For (\ref{eq:IIA}), we find after a few steps using (\ref{eq:piUnten})
\begin{alignat}{3}
(\ref{eq:IIA}) =& \hdots = 2\zeta \int_{\sigma} d^3x ~ \alpha_f\m^e ~ \nabla_g\left( \beta_h\m^g \cdot \text{Tr}\left(\pi_e ~ \pi^f \overset{(\gamma,\beta)}{\pi}\m^h \right) \right) \nonumber \\
=& 2\zeta \int_{\sigma} d^3x ~ \alpha_{fe} \left[ \nabla_g\left( \frac{1}{q} \beta_h\m^g \cdot \text{Tr}\left(\pi^{[e} ~ \pi^{f]} \overset{(\gamma,\beta)}{\pi}\m^h \right) \right) - \left( \nabla_g q^{ea}\right) \frac{1}{q} \beta_h\m^g \cdot \text{Tr}\left(\pi_a ~ \pi^f \overset{(\gamma,\beta)}{\pi}\m^h \right) \right] \approx 0 \nonumber \text{,}
\end{alignat}
which vanishes since the trace $\text{Tr}(abc) := a^I\m_J b^J\m_K c^K\m_I$ of antisymmetric matrices $a$, $b$, $c$ is antisymmetric when exchanging two matrices while $\alpha_{ab}$ is symmetric and $\nabla_a q_{bc} \approx 0$ by construction. The remaining part (\ref{eq:IIB}) can be rewritten as
\begin{alignat}{3}
(\ref{eq:IIB}) =&  - \zeta \int_{\sigma} d^3y ~ \left\{P^{ab}[A_{ab}], \Gamma_{gc}^d(y)\right\} \beta_h\m^g~ \text{Tr}\left(\pi_{d} ~\pi^{c} \overset{(\gamma,\beta)}{\pi}\m^{h}\right) (y) \nonumber \\
=& - \zeta \int_{\sigma} d^3y ~ \left[ \left\{P^{ab}[A_{ab}], q^{de}(y)\right\} q_{di} \Gamma_{gc~e} +  \left\{P^{ab}[A_{ab}], \Gamma_{gc~i}(y)\right\} \right] \frac{1}{q} \beta_h\m^g~ \text{Tr}\left(\pi^{i} ~\pi^{c} \overset{(\gamma,\beta)}{\pi}\m^{h}\right) (y) \nonumber \\
=& - \zeta \int_{\sigma} d^3y ~ \left[ - \left\{P^{ab}[A_{ab}], q_{jk}(y)\right\} \delta_{i}^{(j} \Gamma_{gc}^{k)} +  \left\{P^{ab}[A_{ab}], \partial_c q_{ig}(y)\right\} \right] \frac{1}{q} \beta_h\m^g~ \text{Tr}\left(\pi^{i} ~\pi^{c} \overset{(\gamma,\beta)}{\pi}\m^{h}\right) (y) \nonumber \\
\approx& - \zeta \int_{\sigma} d^3y ~ \left[ A_{ab} \Gamma_{gc}^{b} + A_{ag} \vec{\partial}_c \right] \frac{1}{q} \beta_h\m^g~ \text{Tr}\left(\pi^{a} ~\pi^{c} \overset{(\gamma,\beta)}{\pi}\m^{h}\right)\nonumber \\
=& - \zeta \int_{\sigma} d^3y ~ A_{ab} \nabla_c\left( \frac{1}{q} \beta_h\m^b~ \text{Tr}\left(\pi^{a} ~\pi^{c} \overset{(\gamma,\beta)}{\pi}\m^{h}\right)\right)\nonumber\\
=& \zeta \int_{\sigma} d^3y ~ \left(\nabla_c A_{ab}\right) \frac{1}{4q} B_{ed} \left( q^{b(e} \delta^{d)}_h - q^{cd} \delta_h^b\right)~ \text{Tr}\left(\pi^{a} \overset{(\gamma,\beta)}{\pi}\m^{c} ~\pi^{h} \right) \nonumber\\
=& \zeta \int_{\sigma} d^3y ~ \left(\nabla_c A_{ab}\right) \frac{1}{4q} B^b\m_d ~ \text{Tr}\left(\pi^{a} \overset{(\gamma,\beta)}{\pi}\m^{c} ~\pi^{d} \right) \nonumber \text{.}
\end{alignat}
In the first step, we just reassembled the terms on the left hand side of the Poisson bracket, in the second we used the definition of the Christoffel symbol, in the third the formula for the derivative of the inverse matrix and antisymmetry of the trace in $(i \leftrightarrow c)$. In the fourth line we used the already known brackets of the metric $q_{ab}$ and its conjugate momentum $P^{cd}$. Note that the density weight and index structure is such that the terms in the fourth line can be reassembled in a covariant derivate. In the sixth line the definition of $\beta_h\m^g$ is inserted, we integrated by parts and we used that $\text{Tr}(ab\overset{(\gamma,\beta)}{c}) = \text{Tr}(a\overset{(\gamma,\beta)}{b}c)$ (this trace property can be shown using the definition of the matrices $\overset{(\gamma)}{\mathcal{M}}$). Thus we find that the second summand appearing in the definition of $\beta_h\m^g$ vanishes due to antisymmetry of the trace in the indices $(a \leftrightarrow b)$. If we now restore the antisymmetry in the test fields $(A \leftrightarrow B)$, we obtain
\ba
(\ref{eq:IIB}) &=& \frac{\zeta}{4} \int_{\sigma} d^3y ~ \left[\left(\nabla_c A_{ab}\right) B^b\m_d - \left(\nabla_c B_{ab}\right) A^b\m_d \right] \frac{1}{q}  \text{Tr}\left(\pi^{a} \overset{(\gamma,\beta)}{\pi}\m^{c} ~\pi^{d} \right) \nonumber\\
&\approx& \frac{1}{4 \beta \gamma} \int_{\sigma} d^3y ~  \epsilon^{cda} ~ \nabla_c \left(A_{ab} B^b\m_d \right) \nonumber \\
&=& \frac{1}{4 \beta \gamma} \int_{\sigma} d^3y ~  \partial_c \left( \epsilon^{cda} A_{ab} B^b\m_d \right) = 0 \nonumber \text{,}
\ea 
where we used the simplicity constraint in the second line and then dropped a surface term. We leave the case where $\sigma$ has a boundary for further research. This furnishes the proof of the validity of the formulation. 

\subsection{Linear Simplicity Constraints}
\label{sec:Lin}

As will be demonstrated in \cite{BTTVI}, we can extend the result to the case of a linear simplicity (and normalisation) constraints when introducing additional phase space degrees of freedom $\{N^I, P_J\}$. The theory with linear simplicity constraint has the non-vanishing Poisson brackets (\ref{eq:PoissonBrackets}) as well as
\be
\left\{N^I(x), P_J(y)\right\} = \delta^I_J \delta^3(x-y) \text{,}
\ee
and the constraints are given by
\ba
\mathcal{H}_a &:=& \frac{1}{2} \overset{(\gamma,\beta)}{\pi}\m^{bIJ} \partial_a A_{bIJ} - \frac{1}{2} \partial_b \left( \overset{(\gamma,\beta)}{\pi}\m^{bIJ} A_{aIJ} \right) + P_I \partial_a N^I \text{,} \\
\mathcal{H} &:=& - \frac{s}{8\sqrt{q}} \left( \overset{(\gamma,\beta)}{\pi}\m^{[a|IJ}\overset{(\gamma,\beta)}{\pi}\m^{b]KL} \left(A-\Gamma \right)_{bIJ} \left(A-\Gamma \right)_{aKL}\right) - \sqrt{q} R(\overset{(\gamma,\beta)}{\pi}) \text{,} \\
 G^{IJ} &:=& D_a\overset{(\gamma,\beta)}{\pi}\m^{aIJ} + 2 P^{[I} N^{J]} \text{,} \\
 S^{aI} &:=& \epsilon^{IJKL} N_J \pi^{a}_{KL} \text{,} \\
 \mathcal{N} &:=& N^I N_I - \zeta \text{.}
\ea
Note that for the proof that the ADM Poisson brackets are reproduced on the extended phase space in section \ref{app:Quadratic} we just needed $\bar{G}^{IJ} \approx 0$ and $S^{ab} \approx 0$. Since the solutions to the quadratic and linear simplicity (and normalisation) constraints coincide\footnote{Note that in the considerations in section \ref{app:Quadratic}, we neglected the topological sector when solving the quadratic simplicity constraints.} and on shell, we did not change the rotational parts of the Gau{\ss} constraint, this implies that the ADM brackets are also reproduced in the case at hand if we express $q_{ab}$ and $P^{cd}$ again as given in equations (\ref{eq:Metric}, \ref{eq:MetricConjugateMomentum}). Therefore, we just need to check if the constraint algebra closes. The change in the definition of $G^{IJ}$ and $\mathcal{H}_a$ is needed since the newly introduced constraints $S^{aI}$ and $\mathcal{N}$ have to transform nicely under spatial diffeomorphisms and gauge transformations for the constraint algebra to close. Since the hybrid spin connection (\ref{eq:SpinConnection}) transforms covariantly under spatial diffeomorphisms by inspection and we already noted that it transforms as a connection under gauge transformations, also the Hamilton constraint transforms covariantly under the action of $G^{IJ}$ and $\mathcal{H}_a$. The only non-trivial Poisson bracket is the one of two Hamilton constraints. But since the ADM brackets are reproduced, we know that the result of this Poisson bracket on-shell is the ADM spatial diffeomorphism constraint. Therefore, we just need to show that $\mathcal{H}_a$ reduces correctly. This will be shown in the following section and completes the proof of the validity of this formulation.

\subsection{Solving the Linear Simplicity and Normalisation Constraints}
It is instructive to solve first the simplicity and then the boost Gau{\ss} constraint (``time gauge"), which will in the first step lead to a formulation similar to the one given in \cite{SaHamiltonianAnalysisOf} and then to the formulation in Ashtekar-Barbero variables. We will treat the case with linear simplicity constraints, since in $3+1$ dimensions, the linear constraint has the additional advantage that its only solution is General Relativity, while the quadratic simplicity constraint also allows for the topological solution. 

The solution to the linear simplicity and normalisation constraint is given by $\pi^{aIJ} = 2n^{[I}E^{a|J]}$ (cf. \cite{BTTVI}) and therefore
\be
\overset{(\gamma,\beta)}{\pi}\m^{aIJ} = \frac{1}{\beta}\left(2n^{[I}E^{a|J]} + \frac{1}{\gamma} \epsilon^{IJKL} n_K E_{aL}\right)\text{.}
\ee
For the connection, we make the Ansatz $A_{aIJ} = \Gamma_{aIJ} + \underset{(\gamma,\beta)}{K}\m_{aIJ}$. Using the results of \cite{PeldanActionsForGravity, BTTI, BTTVI}, the symplectic potential becomes
\ba
\frac{1}{2}\overset{(\gamma,\beta)}{\pi}\m^{aIJ} \dot A_{aIJ} + P^I \dot{N_I} &=& \frac{1}{2}\overset{(\gamma,\beta)}{\pi}\m^{aIJ} \dot \Gamma_{aIJ} + \frac{1}{2}\overset{(\gamma,\beta)}{\pi}\m^{aIJ} \dot{\underset{(\gamma,\beta)}{K}}\m_{aIJ} + P^I \dot{N_I} \nonumber \\
&=&\frac{1}{2\beta \gamma} \epsilon^{IJKL} n_K E_{aL} \dot \Gamma_{aIJ} + n^{[I} E^{a|J]} \dot K_{aIJ} + P^I \dot{N_I} \nonumber \\
&=& - \dot{(n^{[I} E^{a|J]})} \left( K_{aIJ} + \frac{1}{2\beta \gamma} \epsilon_{IJ}\m^{KL}  \Gamma_{aKL}\right) + P^I \dot{N_I} \nonumber \\
&\approx& - \dot{E^{aJ}} \left( n^I K_{aIJ} + \frac{1}{2\beta \gamma} \epsilon_{IJ}\m^{KL} n^I \Gamma_{aKL} + n_J P^I E_{aI} \right. \nonumber \\
&\m& \hspace{10mm} \left. - n_J E_a^I E^{bK} \left( \bar{K}_{bIK}  + \frac{1}{2\beta \gamma} \epsilon_{IK}\m^{LM} \Gamma_{bLM}\right)\right) \nonumber \\
&=:& E^{aJ} \dot{A}_{aJ} \text{.}
\ea
In the next step we express the remaining constraints in terms of the new canonical variables. The reduction of the Gau{\ss} constraint yields
\ba
\frac{1}{2} \Lambda_{IJ} G^{IJ} &=& \frac{1}{2} \Lambda_{IJ} D_a\overset{(\gamma,\beta)}{\pi}\m^{aIJ} + \Lambda_{IJ} P^I N^J \nonumber \\
&\approx& \Lambda_{IJ} \underset{(\gamma,\beta)}{K}\m_{a}^{I}\m_{K} \overset{(\gamma,\beta)}{\pi}\m^{aKJ} + \Lambda_{IJ} P^I N^J \nonumber \\
&=& \Lambda_{IJ} K_{a}^{I}\m_{K} \pi^{aKJ} + \Lambda_{IJ} P^I N^J  \nonumber \\
&=& \Lambda_{IJ} K_{a}^{I}\m_{K} \left( n^K E^{aJ} - n^J E^{aK} \right) + \Lambda_{IJ} P^I N^J \nonumber \\
&\approx& - \Lambda_{IJ} E^{aJ} \left(  n_K K_{a}^{KI} - n^I E_{aL} E^{bK} K_{b}^{L}\m_{K} + n^I P^K  E_{aK}\right) \nonumber \\
&=& - \Lambda_{IJ} E^{aJ} \left( A_{a}^{I} - \frac{1}{2\beta \gamma} \left(\epsilon_{M}\m^{IKL} n^M \Gamma_{aKL} - n^I E_a^N E^{bK} \epsilon_{NK}\m^{LM} \Gamma_{bLM}\right) \right) \nonumber \\
&=& \Lambda_{IJ} \left( E^{aI} A_{a}^{J} + \frac{1}{2\beta \gamma} \epsilon^{IJKL} \partial_{a}\left( n_K E^a_L \right) \right) \text{,}
\ea
which after time gauge $n^I = \delta^I_0$ and solution of the boost part of the Gau{\ss} constraint obviously reproduces the SU$(2)$ Gau{\ss} constraint of the Ashtekar-Barbero formulation. The diffeomorphism constraint becomes
\ba
\mathcal{H}_a &=&  \frac{1}{2} \overset{(\gamma,\beta)}{\pi}\m^{bIJ} \partial_a A_{bIJ} - \frac{1}{2} \partial_b \left( \overset{(\gamma,\beta)}{\pi}\m^{bIJ} A_{aIJ} \right) + P_I \partial_a N^I \approx ... \nonumber \\
&\approx& E^{bI} \partial_a A_{bI} - \partial_b\left( E^{bI}A_{aI}\right)\text{.}
\ea
In order to complete the proof of the validity of the formulation with linear simplicity constraints of section \ref{sec:Lin}, we have to show that this constraint becomes the ADM spatial diffeomorphism constraint after solving the Gau{\ss} constraint. Alternatively, we can show that it coincides on-shell with the spatial diffeomorphism constraint of section \ref{app:Quadratic},
\ba
\mathcal{H}_a &=&  -\frac{1}{2} \nabla_b \left( \left( A-\Gamma \right)_{aIJ} \overset{(\gamma,\beta)}{\pi}\m^{bIJ} - \delta_a^b \left( A-\Gamma \right)_{cIJ}\overset{(\gamma,\beta)}{\pi}\m^{cIJ} \right) \approx ... \nonumber \\
&\approx& E^{bI} \partial_a A_{bI} - \partial_b\left( E^{bI}A_{aI}\right) + \frac{1}{2} \Gamma_{aKL} G^{KL} \text{,}
\ea
which it does. For the above calculations, the following results may be helpful:
\ba
\bar{R}_{abIJ} &=& E^{c}_I E_{dJ} R_{abc}\m^d \text{,} \\
 R_{abIJ} n^I E^{bJ} &=& 0 \text{,} \\
 \epsilon^{IJKL} R_{abIJ} n_K E^b_{L} &=& 0 \text{.}
\ea
For the first line, expand an internal vector $\lambda_I$ with $\lambda_I n^I = 0$ into $\lambda_I = e^{a}_I \lambda_a$, $\lambda_a = e_a^I \lambda_I$, use $D_a e_b^J = 0$ and compare $[D_a, D_b]\lambda_J$ with $[D_a, D_b] \lambda_c$. The second line follows from the fact that $\Gamma_{aIJ}$ annihilates $n^I = n^I(E)$ and the third line is a consequence of the first line and the algebraic Bianci identity. Finally, the Hamilton constraint gives
\ba
\mathcal{H} &=& - \frac{s}{8\sqrt{q}} \left( \overset{(\gamma,\beta)}{\pi}\m^{[a|IJ}\overset{(\gamma,\beta)}{\pi}\m^{b]KL} \left(A-\Gamma \right)_{bIJ} \left(A-\Gamma \right)_{aKL}\right) - \sqrt{q} R(\overset{(\gamma,\beta)}{\pi}) \\ 
&\approx& - \frac{s}{2\sqrt{q}} E^{[a|I} E^{b]J} \left(A_{bI} - \frac{1}{2\beta\gamma} \epsilon_{MI}\m^{KL} n^M \Gamma_{bKL}\right) \left(A_{aJ} - \frac{1}{2\beta\gamma} \epsilon_{NJ}\m^{AB} n^N \Gamma_{aAB}\right) - \sqrt{q} R(E) \nonumber \text{.}
\ea
We leave the task of exactly relating this formulation to the similar one given in \cite{SaHamiltonianAnalysisOf} for future research.

\subsection{Time Gauge}
Choosing time gauge $n^I = \delta^I_0 \Leftrightarrow E^{a0} = 0$ always implies solving its second class partner, the boost part $G^{0i} = - E^{ai} A_{a}^0$ of the Gau{\ss} constraint ($i,j,\hdots \in \{1,2,3\}$). In order to obtain expressions which can be easily compared to results in the literature (e.g. \cite{ThiemannModernCanonicalQuantum}), we introduce the rescaled\footnote{The factor of two appears since we want to obtain the Poisson brackets $\left\{A'_{ai}(x), E'\m^{bj}(y)\right\} = \frac{1}{2} \delta_a^b \delta_i^j \delta^3(x-y)$, which brings the formulation closer to the one given in \cite{ThiemannModernCanonicalQuantum}.} variables $A_{ai} \rightarrow A'_{ai} := \frac{\tilde{\gamma}}{2} A_{ai}$ and $E^{bj} \rightarrow E'\m^{bj} := \frac{1}{\tilde \gamma} E^{bj}$, where $\tilde{\gamma} := 2 \zeta \beta \gamma$. The result is
\ba
E^{aI} \dot{A}_{aI} &\rightarrow& 2 E'\m^{ai} \dot{A}'_{ai} \text{,}\\
G^{IJ} &\rightarrow& \frac{1}{4} \epsilon^{kij} G_{ij} =  \partial_a E'\m^{ak} - \epsilon^{kij} A'_{ai} E'\m^a_{j} \text{,} \\
\mathcal{H}_a &\rightarrow& \frac{1}{2} \mathcal{H}_a = E'\m^{bi} \partial_a A'_{bi} - \partial_b\left( E'\m^{bi}A'_{ai}\right) \text{,} \\
\mathcal{H} &\rightarrow& \mathcal{H} = - \frac{s}{2\sqrt{q}} E^{[a|i} E^{b]j} \left(A_{bi} - \frac{1}{\tilde{\gamma}} \epsilon_{i}\m^{kl} \Gamma_{bkl}\right) \left(A_{aj} - \frac{1}{\tilde{\gamma}} \epsilon_{j}\m^{mn}  \Gamma_{amn}\right) - \sqrt{q} R(E) ~~~~~~~~~~~~\\
&\m& \hspace{5mm} \approx - \frac{s}{\sqrt{q}} F'_{abij} E'\m^{ai} E'\m^{bj} + \left(\frac{s}{\tilde{\gamma}^2} - 1\right) \sqrt{q} R(\tilde{\gamma} E') \text{,}
\ea
where terms proportional to the Gau{\ss} constraint have been dropped in the expression for the Hamilton constraint (cf. \cite{ThiemannModernCanonicalQuantum}). At this stage, only the combination of the parameters $\gamma \cdot \beta$ is left and one could ask if one should have worked with one parameter from the beginning. Note that the (quadratic) simplicity constraint implies $ \frac{1}{2} \epsilon_{IJKL} \overset{(\gamma,\beta)}{\pi}\m^{aIJ} \overset{(\gamma,\beta)}{\pi}\m^{bKL} = \frac{2 \zeta\gamma} {\gamma^2 + \zeta} \overset{(\gamma,\beta)}{\pi}\m^{aIJ} \overset{(\gamma,\beta)}{\pi}\m^{b}_{IJ}$ and therefore
\ba
2 \zeta q q^{ab} &=& \pi^{aIJ} \pi^{b}_{IJ} = \left(\frac{\gamma^2}{\gamma^2 - \zeta}\right)^2\left[ \left(1+\frac{\zeta}{\gamma^2}\right)\overset{(\gamma,\beta)}{\pi}\m^{aIJ} \overset{(\gamma,\beta)}{\pi}\m^{b}_{IJ} - \frac{1}{\gamma} \epsilon_{IJKL} \overset{(\gamma,\beta)}{\pi}\m^{aIJ} \overset{(\gamma,\beta)}{\pi}\m^{bKL} \right] \nonumber \\
&\approx& \frac{\gamma^2 \beta^2}{\gamma^2+\zeta}  \overset{(\gamma,\beta)}{\pi}\m^{aIJ} \overset{(\gamma,\beta)}{\pi}\m^{b}_{IJ}\text{.}
\ea
We expect that the square root of this factor, $\frac{\gamma \beta}{\sqrt{\gamma^2+\zeta}}$, will appear in the spectrum of the area operator. We leave the question if the two parameters $\gamma$, $\beta$ appear just in this peculiar combination in the spectra of operators or not, i.e. if one can actually distinguish between $\gamma$ and $\beta$ at the quantum level, for further research.

\subsection{Formulation with two commuting SU$(2)$ Connections}
Note that we could have chosen time gauge before solving the simplicity and normalisation constraints by setting $N^{I} = \delta^I_0$ and solving the boost part of the Gau{\ss} constraint $G^{0i} = D_a \overset{(\gamma,\beta)}{\pi}\m^{ai} - P^i$, where we used the notation $\overset{(\gamma,\beta)}{\pi}\m^{ai} := \overset{(\gamma,\beta)}{\pi}\m^{a0i}$. Furthermore, we define $A_{ai} := A_{a0i}$ and $D_a B^i := \partial_a B^i + A_{aij} B^j$. Then we find
\ba
\frac{1}{2}\overset{(\gamma,\beta)}{\pi}\m^{aIJ} \dot A_{aIJ} + P^I \dot{N_I} &\rightarrow& \frac{1}{2}\overset{(\gamma,\beta)}{\pi}\m^{aij} \dot A_{aij} + \overset{(\gamma,\beta)}{\pi}\m^{ai} \dot A_{ai} 
\ea
and
\ba
G^{IJ} &\rightarrow& G^{ij} = D_a\overset{(\gamma,\beta)}{\pi}\m^{aij}  + 2 \overset{(\gamma,\beta)}{\pi}\m^{a[i}A_a^{j]}  \text{,} \\
S^{aI} &\rightarrow& S^{ai} = \epsilon^{ijk} \overset{(\gamma,\beta)}{\pi}\m^{a}_{jk} - \frac{2\zeta}{\beta} \overset{(\gamma,\beta)}{\pi}\m^{ai} \text{,} \\
\mathcal{H}_a &\rightarrow& \mathcal{H}_a = \frac{1}{2} \overset{(\gamma,\beta)}{\pi}\m^{bij} \partial_a A_{bij} - \frac{1}{2} \partial_b \left( \overset{(\gamma,\beta)}{\pi}\m^{bij} A_{aij} \right) + \overset{(\gamma,\beta)}{\pi}\m^{bi} \partial_a A_{bi} - \partial_b \left( \overset{(\gamma,\beta)}{\pi}\m^{bi} A_{ai} \right) \text{,}~~~~~~~ \\
\mathcal{H} &\rightarrow& \mathcal{H}= - \frac{s}{8\sqrt{q}} \left( \overset{(\gamma,\beta)}{\pi}\m^{[a|ij}\overset{(\gamma,\beta)}{\pi}\m^{b]kl} \left(A-\Gamma \right)_{bij} \left(A-\Gamma \right)_{akl}\right)  \nonumber \\ 
 &\m& \hspace{8mm} - \frac{s}{4\sqrt{q}} \left( \overset{(\gamma,\beta)}{\pi}\m^{[a|i}\overset{(\gamma,\beta)}{\pi}\m^{b]j} A_{bi} A_{aj}\right) - \sqrt{q} R(\overset{(\gamma,\beta)}{\pi}) \text{,}
\ea
where we dropped constants in front of the simplicity constraint and in the Hamilton constraint we neglected terms proportional to the simplicity constraint ($\Gamma_{a0i} \approx 0$). Note that in the case without Barbero-Immirzi parameter, the Simplicity constraint $S^{ai} = \epsilon^{ijk} \pi^{a}_{jk}$ demands the vanishing of $\pi^{aij}$ and therefore there is no physical information left in the conjugate SU$(2)$ connection $A_{aij}$. Here, this is not the case and we obtain a genuine connection formulation of General Relativity. Moreover, the other canonical pair $\left\{ A_{ai}, \overset{(\gamma,\beta)}{\pi}\m^{bj} \right\}$ has the same structure as $\left\{K_{ai}, E^{bj}\right\}$. Then, it follows from the know results when extending the ADM phase space to Ashtekar-Barbero variables (cf., e.g., \cite{ThiemannModernCanonicalQuantum}) that there exists a spin connection $\Gamma'_{aij}$ which annihilates 
$\overset{(\gamma,\beta)}{\pi}\m^{ai}$ and that the transformation $\left\{ A_{ai}, \overset{(\gamma,\beta)}{\pi}\m^{bj} \right\} \rightarrow \left\{A^-_{aij} := \Gamma'_{aij} + \alpha \epsilon_{ijk} A_a^k, ~ E_-\m^{bkl} := \frac{1}{\alpha}  \epsilon^{klm} \overset{(\gamma,\beta)}{\pi}\m^{b}_m \right\}$ for $\alpha \in \mathbb{R}/\{0\}$ is canonical. Defining $A^+_{aij} := A_{aij}$ and $E_+\m^{aij} :=  \overset{(\gamma,\beta)}{\pi}\m^{aij}$, we obtain the symplectic potential 
\ba
&\m& \frac{1}{2} E_+\m^{aij} \dot{A}^+_{aij} + \frac{1}{2} E_-\m^{aij} \dot{A}^-_{aij}
\ea
and constraints
\ba
G^{ij} &=& D^+_a E_+\m^{aij} + D^-_a E_-\m^{aij}  \text{,} \\
\frac{1}{2}\epsilon^{ijk} S^{a}_k &=& E_+\m^{aij} - \frac{\zeta \alpha}{\beta} E_-\m^{aij} \text{,} \\
\mathcal{H}_a &=& \frac{1}{2} E_+\m^{bij} \partial_a A^+_{bij} - \frac{1}{2} \partial_b \left(E_+\m^{bij} A^+_{aij} \right) \nonumber \\
& & +\frac{1}{2} E_-\m^{bij} \partial_a A^-_{bij} - \frac{1}{2} \partial_b \left(E_-\m^{bij} A^-_{aij} \right) - \frac{1}{2} E_-\m^{bij} R'_{abij}(E_-) \text{,} \\
\mathcal{H} &=& - \frac{s}{8\sqrt{q}} \left( E_+\m^{[a|ij} E_+\m^{b]kl} \left(A^+-\Gamma(E_+\m, E_-) \right)_{bij} \left(A^+-\Gamma(E_+\m, E_-) \right)_{akl}\right) \\ 
 &\m&-  \frac{s}{8\sqrt{q}} \left(E_-\m^{[a|ij} E_-\m^{b]kl} \left(A^--\Gamma'(E_-) \right)_{bij} \left(A^- -\Gamma'(E_-) \right)_{akl} \right) - \sqrt{q} R(E_+\m, E_-) \text{,} \nonumber
\ea
where the last term in the spatial diffeomorphism constraint, $E_-\m^{bij} R'_{abij}(E_-)$, vanishes due to the Bianci identity. We made explicit that the spin connection $\Gamma_{aij}$ in the Hamilton constraint does not annihilate $E_+\m^{aij}$ but the physical combination of $E_+\m^{aij}$ and $E_-\m^{aij}$ (i.e. the combination which remains when solving the simplicity constraint). In this formulation we now have two commuting SU$(2)$ connections $A^+_{aij}$ and $A^-_{aij}$, which can be interpreted as the two parts SU$(2)^+$ and SU$(2)^-$ of SO$(4)$. They are, however, not uncorrelated and their momenta are multiples of each other, in complete analogy to the relation $K + \gamma L = 0$ of boost- and rotation generators in the new Spin Foam models.\\
\\
\\
\\
\\
\\
\\
\\
\\
\\

\end{appendix}

\bibliography{pa88pub.bbl}

\end{document}